\newif\ifuseprd
\newif\ifeprint
\newif\ifdatelast

\useprdtrue
\eprinttrue
\datelastfalse

\documentclass[preprint,
tightenlines,%
floats,floatfix,%
prd,eqsecnum,nobibnotes%
,nofootinbib,aps,10pt]{revtex4}
\usepackage{amsmath,amssymb,amsxtra}
\usepackage{graphicx}
\let\oldappendix\appendix
\renewcommand\appendix{\oldappendix%
    \renewcommand\theequation{\thesection.\arabic{equation}}}

\allowdisplaybreaks[3]

\newif\iftoomuchdetail
\toomuchdetailtrue
\toomuchdetailfalse
\newcounter{saveequation}
\newcounter{detailnum}
\newcommand\savetheequation{\theequation}
\newcommand\detailtheequation{%
      $\delta$\Roman{detailnum}:\roman{equation}}

\newcounter{case}

\newenvironment{caselist}{\begin{list}{{\bf Case \Alph{case}:}}{%
   \setlength{\labelwidth}{5em}
   \addtolength{\leftmargin}{\labelwidth}
   \usecounter{case}}}{%
   \end{list}}

\newcommand\abs[1]{\ensuremath{\left\lvert{#1}\right\rvert}}
\newcommand\anti[2]{\ensuremath{\left\{{#1},{#2}\right\}}}
\newcommand\com[2]{\ensuremath{\left[{#1},{#2}\right]}}

\DeclareMathOperator{\Tr}{Tr}
\DeclareMathOperator{\sgn}{sgn}
\newcommand\su{\ensuremath{\mathfrak su}}
\newenvironment{spmatrix}{\left(\begin{smallmatrix}}{\end{smallmatrix}\right)}

\newcommand\mathone{{\rlap{\kern .25em l}1}}
\newcommand\one{{\ifmmode{\text{\mathone}}\else{\mathone}\fi}}

\newcommand\order[1]{{\ensuremath{{\mathcal O}\left({#1}\right)}}}

\newcommand\ppwave{{\em pp}~wave}
\newcommand\ppWave{{\em pp}~Wave}

\makeatletter
\def\@strike{\relax\leavevmode
  \ifmmode
    \expandafter\mathpalette\expandafter\math@strike
  \else
    \expandafter\make@strike
  \fi}
\def\math@strike#1#2{%
  \setbox\z@\hbox{$\m@th#1{#2}$}\fin@strike}
\def\make@strike#1{%
  \setbox\z@\hbox{\color@begingroup#1\color@endgroup}\fin@strike}
\def\fin@strike{%
  \@tempdima\dp\z@
  \@tempdimb\ht\z@
  \lower\@tempdima\hbox{\strike@start}%
  \box\z@
  \raise\@tempdimb\hbox{\strike@end}}
\def\strike@start{\special{ps: %
    currentpoint /starty exch def /startx exch def}}
\def\strike@end{
}
\newcommand\fs{\protect\@strike}

\newcommand\citejournal[4]{{\ifuseprd\else\begingroup\em\fi {#4}%
     \ifuseprd\else\endgroup\fi {\bf {#1}}\ifdatelast, {#3} ({#2})\else%
     \ ({#2}) {#3}\fi}}

\providecommand\plb[3]{{\citejournal{#1}{#2}{#3}{Phys.\ Lett.\ B }}}
\providecommand\npb[3]{{\citejournal{{\ifuseprd\bf\else\em\fi B\/}#1}{#2}{#3}{Nucl.\ Phys.\ }}}
\providecommand\jhep[3]{{\citejournal{#1}{#2}{#3}{J.\ High Energy Phys.\ }}}
\providecommand\npps[3]{{\citejournal{}{#2}{#3}{Nucl.\ Phys. %
     {\normalfont\bf {#1}} Proc.\ Suppl.\ }}}

\providecommand\citeprd[3]{{\citejournal{#1}{#2}{#3}{Phys.\ Rev.\ D }}}

\providecommand\hepth[1]{{\ifuseprd{\eprint{{\ifeprint\tt\fi hep-th/#1}}}%
                \else{\tt hep-th/{#1}}\fi}}

\newcommand\phepth[1]{{\ifuseprd\else\tt\fi [\hepth{#1}]}}
\newcommand\ct[1]{{\ifeprint\ifuseprd{\em{#1}},\else{\sf {#1}},\fi\fi}}
\newcommand\bt[1]{{\em {#1}},}

\newcommand\ben{\begin{equation}}
\newcommand\een{\end{equation}}
\newcommand\bea{\begin{eqnarray}}
\newcommand\eea{\end{eqnarray}}

\newcommand\eg{{\em e.g.\/}}
\newcommand\ie{{\em i.e.\/}}
\newcommand\cf{{\em cf.\/}}
\newcommand\etc{{\em etc.\/}}

\newcommand{\BE}{\begin{eqnarray}}
\newcommand{\EE}{\end{eqnarray}}
\newcommand{\be}{\begin{eqnarray}}
\newcommand{\ee}{\end{eqnarray}}

\newcommand\skipthis[1]{{}}

\newcounter{subtable}[table]
\renewcommand\thesubtable{(\alph{subtable})}
\newlength{\tablelen}
\newcommand\subtable[2][]{\refstepcounter{subtable}%
     \settowidth{\tablelen}{{#2}\thesubtable}
     \begin{tabular}[b]{c} {#2} \\
     {\begin{minipage}{\tablelen}
      \begin{center} \thesubtable\ {#1}%
      \end{center} \end{minipage}} \end{tabular}}

\newlength{\Acclen}
\newlength{\Accheight}
\newcommand\CustomAccent[2]{\settowidth{\Acclen}{\ensuremath{#1}}%
   \settoheight{\Accheight}{\ensuremath{#1}}%
   \addtolength{\Accheight}{.3ex}
   \ensuremath{\text{\rlap{\hspace{.5\Acclen}%
      \settowidth{\Acclen}{\ensuremath{{}^{#2}}}%
      \hspace{-.5\Acclen}%
      \ensuremath{\text{\raise\Accheight\hbox{\scriptsize ${#2}$}}}%
      }}}
    \ensuremath{#1}} 
\newcommand\Plus[1]{\CustomAccent{#1}{+}}
\accentedsymbol{\Sp}{\ensuremath{\Plus{\boldsymbol{S}}}}
\accentedsymbol{\Yp}{\ensuremath{\Plus{\boldsymbol{Y}}}}

\allowdisplaybreaks[4]

\begin{document}

\setlength{\baselineskip}{1.2\baselineskip} 

\title{\vspace*{\fill}{
\Large Dynamics of Antimembranes in the
Maximally Supersymmetric
Eleven-Dimensional \ppWave
}}

\author{Jeremy Michelson}\email{jeremy@pa,uky,edu}
\author{Xinkai Wu}\email{xinkaiwu@pa,uky,edu}
\affiliation{Department of Physics and Astronomy \\
       University of Kentucky \\
       Lexington, KY \ 40506 \\ U.S.A.
\vspace*{3\baselineskip} 
}

\begin{abstract}
\vspace*{\baselineskip}
We study a spherical antimembrane in the eleven dimensional \ppwave.
In this background, a single antimembrane breaks all the
supersymmetries because its dipole is misaligned with the background
flux. Using the BMN matrix theory we compute the one-loop potential
for the antimembrane. Then we put the antimembrane in the field
produced by a source spherical membrane and compute the
velocity-dependent part of the interaction between them on both the
supergravity side and the BMN matrix theory side. Despite the
aforementioned nonsupersymmetry of the antimembrane, it is found
that the results on the two sides completely agree.
\vspace*{\fill}
\end{abstract}

\preprint{\parbox[t]{10em}{\begin{flushright}
UK/05-10 \\ {\tt hep-th/0509017}\end{flushright}}}

\maketitle

\ifeprint
\tableofcontents
\fi

\section{Introduction}\label{section: introduction}
Banks, Fischler, Shenker and Susskind proposed \cite{Banks:1996vh}
that M-theory should be described nonperturbatively by what is now
known as the BFSS matrix theory. It has passed many tests.  For
example, it reproduces eleven dimensional supergravity computations
such as interactions between gravitons and other objects~\cite{Banks:1996vh,%
b1,b2,b3,b4,b5,b6,b7,b8}.
However, the study of
BFSS matrix theory is not very easy because the flat directions in
its potential result in a continuous spectrum. The situation is improved
in the maximally supersymmetric eleven dimensional \ppwave\
background for which \cite{Berenstein:2002jq} proposed the BMN matrix
theory. The BMN theory is a mass deformation of the BFSS theory.%
\footnote{For other Matrix theories in
non-flat backgrounds, see \eg~\cite{TRCB}. However, in~\cite{TRCB}
the approximation of a weakly curved background was
made, whereas the BMN matrix theory is an exact one.}
The mass
deformation (parameterized by $\mu$, the
strength of the background four-form)
lifts the flat directions in the potential, giving a
discrete spectrum with the vacua being concentric spherical membrane
configurations. It is then natural to investigate in BMN matrix
theory the dynamics of membranes \cite{Shin:2003np,Shin:2004az},
as well as gauge/gravity dualities
\cite{Lee:2003kf,Lee:2004kv,Bak:2005ef}.

In \cite{Lee:2003kf,Lee:2004kv} the interactions between gravitons,
and between spherical membranes, in the \ppwave%
\footnote{In this paper, ``the'' \ppwave\
refers to the maximally supersymmetric eleven dimensional one.\cite{kg}}
were investigated, and it
was shown that one-loop computations on the matrix theory side
properly reproduce the results on the supergravity side. In this
paper, we continue our work along this line and consider the
dynamics of a spherical antimembrane in the \ppwave\ background.

Of course, in the \ppwave\ the spherical branes
that we call membranes and antimembranes have no net membrane charge.
The distinction between ``membranes'' and ``antimembranes''
is in their nonzero dipoles, which have opposite sign between the two.
Upon taking the flat space limit, one looks at small
regions, \eg\ near the north poles of the spheres, where there is a local
concentration of positive charge for the ``membrane'' and local concentration
of negative charge for the ``antimembrane''; these corresponds to the usual
membrane and antimembrane in flat space.

In flat space, an infinitely extending flat antimembrane is half BPS,
just as a membrane is. In the \ppwave, the nonvanishing background
four-form field strength changes the situation. A spherical membrane
(centered at the origin and of an appropriate size) is half BPS,
while a spherical antimembrane breaks all the supersymmetry. This is
because whereas the dipole of the membrane is ``aligned'' with the
background flux, that of the antimembrane is ``antialigned'' with it.

Using the BMN matrix theory we compute the one-loop potential for
the antimembrane. Then we put the antimembrane in the field produced
by a source spherical membrane (on top of the \ppwave\ background of
course) and compute the velocity-dependent part of the interaction
between them on both the supergravity side and the BMN matrix theory
side. Although the antimembrane breaks all the supersymmetries,
complete agreement is
found between the results on the two sides.

In flat space, \cite{Aharony:1996bh,Lifschytz:1996rw}
considered the interaction of a membrane-antimembrane pair (in the
IIA language, a D2-$\overline{\text{D2}}$ pair with a large number $N$ of
D0's bound to each) and found that the results of computations in
BFSS matrix theory and
supergravity agree. In flat space, the brane and the
antibrane are individually half BPS although the pair is
nonsupersymmetric. Furthermore it can be argued
\cite{Lifschytz:1996rw} that, through a series of T and S dualities,
the D2-$\overline{\text{D2}}$ system (with $N$ D0 branes bound to each) can be
mapped into two clusters of D0-branes moving at a small relative
transverse velocity $v\sim \frac{1}{N}$ and is therefore approximately
supersymmetric for large $N$; this D0 system is
known~\cite{Banks:1996vh} to exhibit agreement between matrix theory
and supergravity results. The work presented in this paper can be
regarded as the generalization of~\cite{Aharony:1996bh,Lifschytz:1996rw}
to the \ppwave\ in some sense, although in the \ppwave\
background there is a big difference because, as we have pointed out, the
 antimembrane itself breaks all the
supersymmetry. In~\cite{Lee:2003kf} it was shown that BMN matrix theory
and supergravity agree on the interaction between two gravitons in the
\ppwave, and one might ask whether, by some analog of
the duality transformations in flat space, the membrane-antimembrane
pair in the \ppwave\ can also be transformed into two clusters of 
D0-branes---more precisely, two clusters of
M-gravitational waves\footnote{We note the curious fact for pp waves that the compactification can break
precisely those supersymmetries preserved by the momentum along the circle
so that D0 branes effectively break all the supersymmetry of the IIA
background, although there is supersymmetry in 11-dimensions.  This
was first noticed in the context of the 26 supercharge \ppwave in
\cite{Michelson:2002ps}. But at the end of the day, we are not interested
in the compact theory, but only use the language of D-branes for
convenience.}---moving at small relative speed,
which would therefore be approximately supersymmetric, as in flat space.
The complete agreement found in this paper for the
velocity-dependent part of the interaction certainly suggests that
this could indeed be the case. However, we still have to compute the
velocity-independent part of the interaction to get the complete story.
Moreover, in flat space, the spatial world
volumes of the D2-branes are tori and one can perform T-duality; for
the \ppwave, the branes are spheres, and at best we do not
understand how to carry out T-duality. Hence, whether such a duality
transformation exists in the \ppwave\ is not completely clear to us at
this stage.

The paper is organized as follows. In Section \ref{section: intro
antimembrane} we identify the antimembrane configuration in both
supergravity and BMN matrix theory. In Section \ref{section: matrix
theory} we compute the one-loop effective potentials on the BMN
matrix theory side, first computing the potential for the
antimembrane by itself, then computing its interaction with the
membrane. In Section \ref{section: sugra Llc} we compute the
membrane-antimembrane interaction on the supergravity side and
compare it with the matrix theory result. We end with a Discussion.
The technical details of diagonalizing the fluctuating modes are
given in the Appendices.

\section{Antimembranes in the eleven dimensional \ppwave}\label{section: intro
antimembrane}
In eleven dimensional supergravity, the
antimembrane's lightcone Lagrangian can be
obtained from that of the membrane by replacing $A_{\mu\nu\rho}$ in
the latter with $-A_{\mu\nu\rho}$. The Lagrangian
density
of a membrane in a general background is%
\footnote{We use indices $i,j,k,\dots = 1,2,3$;
$a,b,c,\dots = 4,\dots,9$; and $I,J,K,\dots = 1,\dots,9$,
unless otherwise stated.
Indices $\mu,\nu,\dots=+,-,1,\dots,9$ are
11-dimensional curved-space indices.}
\BE
{\mathcal L} =
-T\left[\sqrt{-\det(g_{ij})}-\frac{1}{6}\epsilon^{ijk}A_{\mu\nu\rho}
\partial_iX^\mu\partial_jX^\nu\partial_kX^\rho\right],
\EE
and the Lagrangian density of an antimembrane (distinguished from that of a membrane
by use of an overbar) is obtained by flipping
the sign of the Wess-Zumino term, so that
\BE
\bar{{\mathcal L}}=
-T\left[\sqrt{-\det(g_{ij})}+\frac{1}{6}\epsilon^{ijk}A_{\mu\nu\rho}
\partial_iX^\mu\partial_jX^\nu\partial_kX^\rho\right].
\EE

The lightcone Lagrangian density, ${\mathcal
L}_\text{l.c.}(X^A,\dot{X}^A,\partial_rX^A;X^-,\Pi_-,\partial_rX^-)$,
is obtained through Legendre transformation of the $x^-$ degree of freedom.
As
for the untransformed Lagrangian, we flip the sign of
$A_{\mu\nu\rho}$ in ${\mathcal L}_\text{l.c.}$ to get the lightcone Lagrangian
for the antimembrane. For the \ppwave, the only nonvanishing component
of the three-form is
\BE
A_{+ij}=\frac{\mu}{3}\epsilon_{ijk}x^k,
\EE
so equivalently we can replace $\epsilon_{ijk}$ with
$-\epsilon_{ijk}$. Let us consider a spherical antimembrane with
$X^iX^i={r_0'}^2$, $X^a=0$, and M-momentum density
$\Pi_-=\sin\theta\tilde{p}^+$.
After integrating over the sphere,
its lightcone Lagrangian is,
\begin{equation}\label{eqn: Stree sugra}
\begin{split}
\bar{L}_\text{l.c.}&=
4\pi\left[\frac{\tilde{p}^+}{2}\left(\frac{dr_0'}{dt}\right)^2
-\frac{\tilde{p}^+}{18}\mu^2\left(r_0'\right)^2-\frac{T^2}{2\tilde{p}^+}\left(r_0'\right)^4
-T\frac{\mu}{3}\left(r_0'\right)^3\right], \\
&=4\pi\frac{\mu^4(\tilde{p}^+)^3}{18T^2}\left[\left(\frac{1}{\mu}\frac{d\eta}{
dt}\right)^2-\frac{1}{9}\eta^2-\frac{1}{9}\eta^4-\frac{2}{9}\eta^3\right],
\qquad \eta \equiv \frac{3T}{\mu\tilde{p}^+} r_0'.
\end{split}
\end{equation}
Note that the potential $\bar{V}(\eta)\sim \left(
\frac{1}{9}\eta^2+\frac{1}{9}\eta^4+\frac{2}{9}\eta^3\right)=\frac{1}{9}\eta^2(\eta+1)^2$
is a monotonically increasing function with its only minimum being
at $\eta=0$. (Contrast this with the membrane potential $V(\eta)$,
for which the $\eta^3$ term has the
opposite sign, and thus a local minimum at
$\eta=1$.)
Because of the ``wrong'' sign of the dipole it carries,
an antimembrane of constant size does not solve the equation of
motion and is therefore an unstable configuration. This is in
contrast to flat space, for which a single antimembrane is half BPS
and thus stable.

In the BMN matrix theory, the above spherical antimembrane
configuration is given by replacing $X^i$ with $-X^i$ in the usual
fuzzy-sphere solution. Let us see why this is so. The action of the
matrix theory in a generic weakly curved eleven dimensional background
\cite{TRCB,Taylor:1999gq} contains the following term describing the
coupling of the matrix theory object given by $X^i$ with the
background three-form, up to an overall numerical factor
\BE
S_1= J^{MNP(i_1\dots i_n)}(X^i)\partial_{i_1}\cdots\partial_{i_n}A_{MNP}(0),
\EE
($M,N,P=0,1,\dots,10$, $i,i_1,\dots,i_n=1,\dots,9$) where
$J^{MNP(i_1\dots i_n)}(X^i)$ is (the moment of) the three-form
current, and $\partial_{i_1}\cdots\partial_{i_n}A_{MNP}(0)$ is (the
derivative of) the background three-form, evaluated at the origin.
For the \ppwave, this gives
\BE
S_1=
J^{+ij(i_1)}\left[\partial_{i_1}\left(\frac{\mu}{3}\epsilon_{ijk}x^k\right)\right](0)
=J^{+ij(k)}\frac{\mu}{3}\epsilon_{ijk}
=-\frac{\mu}{18R}\Tr\left(i\com{X^i}{X^j}X^k\right)\epsilon_{ijk},
\EE
where in the last line we've used the fact that \cite{Taylor:1999gq}
\BE
J^{+ij(k)}=-\frac{1}{6R}\Tr\left(i\com{X^i}{X^j}X^k\right).
\EE
As one
can see $S_1$ is nothing but the Myers term in the BMN matrix
theory. Now sending $X^i$ to $-X^i$ flips the sign of $S_1$, which
corresponds to the sign-flip of the Wess-Zumino term on the
supergravity side we mentioned earlier. Hence if $X^i$ represents a
spherical membrane, $\bar{X}^i\equiv -X^i$ will represent a
spherical antimembrane.

\section{Computation in the BMN Matrix Theory}\label{section: matrix
theory}
\subsection{Tree Level}
The BMN matrix theory action is \cite{Berenstein:2002jq}
\begin{equation}
\begin{split}
\label{Maction}
{\cal S} =\int dt \Tr\Bigg\{&
\sum_{I=1}^9\frac{1}{2R}(D_t X^I)^2
+ i \psi^TD_t \psi
+ \frac{(M^3R)^2}{4R}\sum_{I,J=1}^9\com{X^I}{X^J}^2 \\
& - (M^3R) \sum_{I=1}^9  \Psi^\dagger \gamma^A \com{\Psi}{X^I} +
\frac{1}{2R}\left[-(\frac{\mu}{3})^2\sum_{i=1}^3(X^i)^2
-(\frac{\mu}{6})^2\sum_{a=4}^9(X^a)^2\right]  \\
& - i \frac{\mu}{4} \Psi^\dagger \gamma_{123} \Psi
- i \frac{(M^3 R)\mu}{3 R}\sum_{i,j,k=1}^3 \epsilon_{ijk} X^i X^j X^k
\Bigg\},
\end{split}
\end{equation}
where $D_t X^I=\partial_tX^I-i\com{X_0}{X^I}$, $M$ is the eleven dimensional
Planck mass, and $R$
is the radius of the M-circle. We also define the parameter
$\alpha=\frac{1}{M^3R}$. In what follows we set both $M$ and $R$ to
one (and hence $\alpha=1$), which can be easily restored later. The
background field configuration is
\BE
B^{I} = \begin{pmatrix} B^I_{(1)}&0\\0&B^I_{(2)} \end{pmatrix},
\EE
where \BE B^i_{(1)}=\frac{\mu}{3}J^i_{(1)},\ \ B^a_{(1)}=0\cdot
I_{N_1\times N_1},\EE represents a spherical membrane sitting at the
origin of all the transverse directions of \ppwave, and \BE
B^i_{(2)}=-\frac{\mu}{3}J^i_{(2)},\ \ B^a_{(2)}=x^a(t)\cdot
I_{N_2\times N_2},\EE represents an antimembrane sitting at the
origin of the $1,2,3$ directions and moving along the trajectory
$x^a(t)$ in the $4,\dots,9$ directions, with the extra minus sign in
$B^i_{(2)}$ appropriate for an antimembrane as explained earlier.
$J^i_{(s)},s=1,2$ is an $N_s\times N_s$ dimensional irreducible
representation of \su(2)
with $\com{J^i_{(s)}}{J^j_{(s)}}=i\epsilon^{ijk}J^k_{(s)}$. The
background values for the gauge field $A$ and the fermions all
vanish. Recall that the Casimir of the $N_s$-dimensional 
irreducible representation
of \su(2)
is given by
$J^i_{(s)}J^i_{(s)}
=\frac{N_s^2-1}{4}\cdot I_{N_s\times N_s}$; hence
the radius of the membrane is
$r_0=\sqrt{\frac{\Tr\left(B^i_{(1)}B^i_{(1)}\right)}{N_1}}
=\frac{\mu}{6}\sqrt{N_1^2-1}$
(which is approximately $\frac{\mu}{6}N_1$, for large $N_1$, or
$\frac{\alpha\mu}{6}N_1$ upon restoring $\alpha$),
and that of the antimembrane is
$r'_0=\sqrt{\frac{\Tr\left(B^i_{(2)}B^i_{(2)}\right)}{N_2}}=\frac{\mu}{6}\sqrt{N_2^2-1}$.

Plugging the above background configuration $B^I$ into
(\ref{Maction}), we find that contributions from $B^I_{(1)}$ cancel
out as expected (since a lone spherical membrane is
supersymmetric) and what is left comes purely from the antimembrane,
\begin{equation}\label{eqn: Stree matrix}
\begin{split}
S_\text{tree}&=\int dt\,
N_2\left\{\frac{1}{2}\dot{x}^a\dot{x}^a-2\left(\frac{\mu}{3}\right)^4\frac{\Tr
J^k_{(2)}J^k_{(2)}}{N_2}
-\frac{1}{2}\left(\frac{\mu}{6}\right)^2x^ax^a\right\},\\
&=\int dt\
N_2\left\{\frac{1}{2}\dot{x}^a\dot{x}^a-2\left(\frac{\mu}{3}\right)^4\frac{N_2^2-1}{4}
-\frac{1}{2}\left(\frac{\mu}{6}\right)^2x^ax^a\right\},
\end{split}
\end{equation}
with the subscript ``tree'' denoting that this is the tree-level
action. The second term, \ie\ the constant potential term in the
above $S_\text{tree}$, stems from the fact that the antimembrane with
a constant radius $r'_0$ is not supersymmetric and does not satisfy
the equation of motion. In fact, recalling that the total M-momentum
carried by the antimembrane is $N_2$, which is equal to $\int
\Pi_-=4\pi\tilde{p}^+$, and also noting that $T=\frac{1}{2\pi}$
(recall $T=\frac{M^3}{2\pi}$), we see that the constant term is equal to the
lightcone Lagrangian (\ref{eqn: Stree sugra})
in the large $N_2$ limit
(upon setting
$r'_0=\frac{\mu\tilde{p}^+}{3T}=\frac{\mu N_2}{6}$, \ie, setting
$\eta=1$).
But we've already commented
after eqn.~(\ref{eqn: Stree sugra}) that $\eta=1$ does not solve the
antimembrane equation of motion.%
\footnote{To get an antimembrane of
radius other than $\frac{\mu N_2}{6}$, \ie\ with $\eta\neq 1$, one
takes $B^i_{(2)}=-h\frac{\mu}{3}J^i_{(2)}$ with $h$ being a pure
number which then gives $r_0'=h\frac{\mu N_2}{6}$, \ie\ $\eta=h$.
Similarly for the membrane.}

Although an antimembrane with a constant radius does not satisfy the
equation of motion, it provides an off-shell background field
configuration whose effective potential can be computed. This is
just as for ordinary field theories---indeed the BMN matrix theory
is just an ordinary quantum mechanical (field) theory. More examples in which
off-shell background field configurations in the BMN matrix theory
were considered, in order to test gauge-gravity duality, can be
found in~\cite{Lee:2003kf,Lee:2004kv}, in which the probe
graviton/membrane was allowed to follow an arbitrary off-shell
trajectory and complete agreement on the two sides of the duality
was found.

\subsection{One-Loop}
The one-loop potential is given by the Coleman-Weinberg formula
(also called the ``sum-over-mass formula''), \BE\label{eqn: sum over
mass} V_\text{eff}^\text{one-loop}=-\frac{1}{2}\left( \sum
m_\text{boson}-\sum m_\text{fermion} -\sum m_\text{ghost}\right),
\EE where the ghosts arise from the standard gauge fixing of the
background field method. (For explicit details in the context of the
\ppwave, see~\cite{Lee:2004kv,Lee:2003kf,Shin:2004az}.) Upon writing
the fluctuations as, \BE X^I=B^I+Y^I, \EE and rescaling
\begin{align}\label{eqn: rescaling}
A&\to\mu^{-1/2}A, & Y^I&\to\mu^{-1/2}Y^I, & C&\to\mu^{-1/2}C,&
\bar{C}&\to\mu^{-1/2}\bar{C},& B^I&\to\mu B^I,& t&\to\mu^{-1}t,
\end{align}
the part of the action that is quadratic in the fluctuating fields no
longer contains $\mu$ explicitly and is given by
\begin{multline}\label{eqn: S2}
S_2=\int dt\Tr \Bigg\{
\frac{1}{2}\left(\dot{Y}^I\right)^2-2i\dot{B}^I[A,Y^I]+\frac{1}{2}\left(
[B^I,Y^J]\right)^2+[B^I,B^J][Y^I,Y^J]-i\epsilon^{ijk}B^iY^jY^k\\
-\frac{1}{2}\left(\frac{1}{3}\right)^2\left(Y^i\right)^2
-\frac{1}{2}\left(\frac{1}{6}\right)^2\left(Y^a\right)^2+i\Psi^\dagger\dot{\Psi}
-\Psi^\dagger\gamma^I[\Psi,B^I]-i\frac{1}{4}\Psi^\dagger\gamma^{123}\Psi\\
-\frac{1}{2}\left(\dot{A}\right)^2-\frac{1}{2}\left([B^I,A]\right)^2
+\dot{\bar{C}}\dot{C}+[B^I,\bar{C}][B^I,C]\Bigg\},
\end{multline}
where
the $\gamma^I$'s are $16\times 16$ real and symmetric $SO(9)$ gamma
matrices.  Write the flucuating fields in block form
\begin{equation}
\begin{aligned}
A&=\begin{pmatrix} Z^0_{(1)}&\Phi^0 \\ {\Phi^0}^\dagger &Z^0_{(2)}
\end{pmatrix}, &
Y^I&=\begin{pmatrix} Z^I_{(1)}&\Phi^I\\ {\Phi^I}^\dagger&
                                Z^I_{(2)}\end{pmatrix}, &
\Psi&=\begin{pmatrix}\Psi_{(1)}&\chi\\
                    \chi^\dagger & \Psi_{(2)}\end{pmatrix}, \\
C&=\begin{pmatrix} C_{(1)}&C\\C^\dagger &C_{(2)}\end{pmatrix}, &
\bar{C}&=\begin{pmatrix} \bar{C}_{(1)}&\bar{C}\\\bar{C}^\dagger &\bar{C}_{(2)}
         \end{pmatrix}.
\end{aligned}
\end{equation}

It is easy to see that
the contributions to $S_2$ from the bosons, fermions, and ghosts
separate, \BE S_2=S_\text{boson}+S_\text{fermion}+S_\text{ghost}.
\EE Furthermore, one can see that the contribution from the diagonal
fluctuations $Z^0_{(s)}, Z^I_{(s)}$, \etc\ (which are $N_s\times N_s$
matrices, $s=1,2$) and that from the off-diagonal fluctuations
$\Phi^0,{\Phi^0}^\dagger$, \etc\ (which are $N_1\times N_2$ or $N_2\times
N_1$ matrices) also separate (using the subscript ``d'' to denote
``diagonal'' and ``o.d.'' to denote ``off-diagonal'')
\begin{equation}
\begin{gathered}
S_\text{boson}=(S_\text{boson})_\text{d}+(S_\text{boson})_\text{o.d.}, \qquad
S_\text{fermion}=(S_\text{fermion})_\text{d}+(S_\text{fermion})_\text{o.d.},\\
S_\text{ghost}=(S_\text{ghost})_\text{d}+(S_\text{ghost})_\text{o.d.}
\end{gathered}
\end{equation}

\subsubsection{Diagonal Fluctuations}\label{subsubsection diagonal}
Below we shall first look at the mass spectrum of the diagonal
fluctuations. Again, one sees that the contributions from the
membrane block and that from the antimembrane block separate, \ie\
\begin{align}
(S_\text{boson})_\text{d}&=\sum_{s=1,2}(S_\text{boson})_{\text{d}(s)}, &
(S_\text{fermion})_\text{d}&=\sum_{s=1,2}(S_\text{fermion})_{\text{d}(s)}, &
(S_\text{ghost})_\text{d}&=\sum_{s=1,2}(S_\text{ghost})_{\text{d}(s)}.
\end{align}
Recall that the background configuration after rescaling by $\mu$ is
given by \BE B^i_{(s)}=\eta_s\frac{1}{3}J^i_{(s)},\quad s=1,2, \EE
with $\eta_1=1$ and $\eta_2=-1$. It is straightforward to show that
\begin{multline} \label{Sboson:diag1}
(S_\text{boson})_{\text{d}(s)}=\Tr\Bigg\{
\frac{1}{2}(\dot{Z}^I_{(s)})^2-\frac{1}{2}(\dot{Z}^0_{(s)})^2
+\frac{1}{2}\left(\frac{1}{3}\right)^2\com{J^i_{(s)}}{Z^J_{(s)}}^2
+\left(\frac{1}{3}\right)^2i\epsilon^{ijk}\com{J^k_{(s)}}{Z^i_{(s)}}Z^j_{(s)}\\
-\frac{1}{2}i\epsilon^{ijk}\eta_s\frac{1}{3}\com{J^i_{(s)}}{Z^j_{(s)}}Z^k_{(s)}
-\left(\frac{1}{3}\right)^2\frac{1}{2}(Z^i_{(s)})^2
-\left(\frac{1}{6}\right)^2\frac{1}{2}(Z^a_{(s)})^2-\frac{1}{2}\left(\frac{1}{3}\right)^2\com{J^i_{(s)}}{Z^0_{(s)}}^2
\Bigg\}.
\end{multline}

Using the fact
\begin{equation}
\Tr\left\{
\left(\frac{1}{3}\right)^2\epsilon^{ijk}\epsilon^{ilm}
      \com{J^j_{(s)}}{Z^k_{(s)}}\com{J^l_{(s)}}{Z^m_{(s)}}\right\}
=\Tr\left\{\left(\frac{1}{3}\right)^2\left(
\com{J^j_{(s)}}{Z^k_{(s)}}^2-\com{J^i_{(s)}}{Z^i_{(s)}}^2
+i\epsilon^{ijk}Z^i_{(s)}\com{J^j_{(s)}}{Z^k_{(s)}}\right)\right\},
\end{equation}
one can rewrite~\eqref{Sboson:diag1} as
\begin{equation}\label{eqn: SBd}
\begin{split}
(S_\text{boson})_{\text{d}(s)}=\frac{1}{2}\Tr\Bigg\{&
-(\dot{Z}^0_{(s)})^2
-\left(\frac{1}{3}\right)^2\com{J^i_{(s)}}{Z^0_{(s)}}^2
+(\dot{Z}^i_{(s)})^2
-\left(\frac{1}{3}\right)^2\left(Z^i_{(s)}
  +i\epsilon^{ijk}\eta_s\com{J^j_{(s)}}{Z^k_{(s)}}\right)^2\\
&+\left(\frac{1}{3}\right)^2\com{J^i_{(s)}}{Z^i_{(s)}}^2
+(\dot{Z}^a_{(s)})^2
-\left(\frac{1}{3}\right)^2\left(\frac{1}{4}(Z^a_{(s)})^2
   -\com{J^i_{(s)}}{Z^a_{(s)}}^2\right)\\
&+(1-\eta_s)\left(\frac{1}{3}\right)^2
  i\epsilon^{ijk}Z^i_{(s)}\com{J^j_{(s)}}{Z^k_{(s)}}\Bigg\}.
\end{split}
\end{equation}
All but the last line of eqn.~(\ref{eqn: SBd}) can be obtained by
replacing $J^i_{(s)}$ in equation~(5.2) of~\cite{Shin:2003np} with
$\eta_s J^i_{(s)}$. For $\eta_s = 1$, the last line vanishes, of
course, and one finds that the membrane configuration satisfies the
equations of motion with zero energy---\ie\ it is BPS.  For $\eta_s
= -1$, the last line explicitly demonstrates the absence of
supersymmetry for the antimembrane configuration.

Now we want to diagonalize $(S_\text{boson})_{\text{d}(s)}$. We use the
$N_s\times N_s$ matrix spherical harmonics $Y^{(s)}_{jm}$
($j=0,\dots,N_s-1$; $m=-j,\dots,j$) to expand the fields, \eg\
\BE
Z^0_{(s)}=\sum_{j=0}^{N_s-1}\sum_{m=-j}^j Z^0_{(s)jm}Y^{(s)}_{jm}.
\EE
As $Z^0_{(s)}$ is not coupled to other fields, finding its mass is
trivial. By noticing that
\BE
\Tr(\dot{Z}^0_{(s)})^2=\sum_{j=0}^{N_s-1}\sum_{m=-j}^j
N_s\left\lvert \dot{Z}^0_{(s)jm} \right\lvert^2, \qquad
\Tr(\com{J^i_{(s)}}{Z^0_{(s)}})^2=-\sum_{j=0}^{N_s-1}\sum_{m=-j}^j
j(j+1)N_s\left\lvert Z^0_{(s)jm} \right\lvert^2,
\EE
one finds the
mass of $Z^0_{(s)jm}$ to be
\BE\frac{1}{3}\sqrt{j(j+1)}.\EE
Similarly
$Z^a_{(s)}$ is not coupled to other fields and one finds the mass of
$Z^a_{(s)jm}$ to be
\BE
\frac{1}{3}\sqrt{\frac{1}{4}+j(j+1)}=\frac{1}{3}\left(j+\frac{1}{2}\right).
\EE
The $Z^i_{(s)}$'s are coupled and finding their masses requires
slightly more work. We relegate the details to
Appendix~\ref{sec:Zmass}, where the masses and degeneracies for
$s=1$ are given in eqn.~(\ref{eqn: mass 1}), (\ref{eqn: deg 1}),
while the masses and degeneracies for $s=2$ are given in
eqns.~(\ref{eqn: mass 2}) and ~(\ref{eqn: deg 1}).
This is summarized in Table~\ref{table:diag}\ref{table:Zi:diag}.

It is worth pointing out that there are twenty one tachyon modes in
the antimembrane spectrum (\ref{eqn: mass 2}), corresponding to mass
$\frac{1}{3}\sqrt{j^2-4j+1}$ for $j=1,2,3$, with the masses-squared
being $-2\left(\frac{1}{3}\right)^2$,
$-3\left(\frac{1}{3}\right)^2$, $-2\left(\frac{1}{3}\right)^2$, and
the degeneracies being $2j+3=5,7,9$ respectively. These correspond
to instabilities in the fluctuations of $X^{1,2,3}$ and will cause
the antimembrane to decay.

Next we look at the contribution from diagonal fluctuations of the
fermion. This part of the action (\ref{eqn: S2}) is given by
\begin{equation}\label{eqn: SFd}
\begin{split}
(S_\text{fermion})_{\text{d}(s)}&=\Tr\left(
i\Psi^\dagger_{(s)}\dot{\Psi}_{(s)}
-\frac{\eta_s}{3}\Psi^\dagger_{(s)}\gamma^i[\Psi_{(s)},J^i_{(s)}]
-i\frac{1}{4}\Psi^\dagger_{(s)}\gamma^{123}\Psi_{(s)}\right),\\
&=\Tr\left(i\psi^{\dagger A\alpha}_{(s)}\dot{\psi}_{(s)A\alpha}
+\frac{\eta_s}{3}\psi^{\dagger A\alpha}_{(s)}(\sigma^i)_\alpha^{\
\beta}[\psi_{(s)A\beta},J^i_{(s)}]-\frac{1}{4}\psi^{\dagger A\alpha}_{(s)}
\psi_{(s)A\alpha}\right),
\end{split}
\end{equation}
where in the last line we have written the spinor in the
SU(2)$\times$SU(4) form $\psi_{A\alpha}$ with $\alpha$ being
SU(2) index and $A$ being SU(4) index, and $\sigma^i$ is the
standard Pauli matrix. One immediately sees that for solutions of
the eigenvalue problem
\BE
(\sigma^i)_\alpha{^{\beta}}\com{J^i_{(s)}}{\psi_{(s)A\beta}}
=\lambda\psi_{(s)A\alpha},
\EE
the
action is diagonalized with the mass of $\psi_{(s)A\alpha}$ given by
$\left\lvert\frac{\eta_s}{3}\lambda+\frac{1}{4}\right\rvert$. This
eigenvalue problem is solved by the
matrix spinor spherical harmonics~\cite{Das:2003yq}
(see also~\cite{dsv1,Shin:2003np})
so we just quote the result,
\begin{equation}
\begin{aligned}
\lambda&=j, &\text{with } j&=0,\dots,N_s-1, &  m&=-j-1,\dots,j, \\
\lambda&=-j-1,& \text{with } j&=1,\dots,N_s-1,& m&=-j,\dots,j-1,
\end{aligned}
\end{equation}
 which gives the masses summarized in
Table~\ref{table:diag}\ref{table:psi:diag}.

\begin{table}[tb]
\begin{center}
\begin{tabular}{cc}
\subtable[$Z^i$\label{table:Zi:diag}]{
\begin{tabular}{|c|c|r@{$\leq j \leq$}l|c|}
\hline
& mass & \multicolumn{2}{c|}{range of $j$} & degeneracy \\
\hline
& $\frac{1}{3} j$ & $1$ & $N_s - 1$ & $2j-1$\\
\cline{2-5}
$\eta_s = 1$ & $\frac{1}{3}\sqrt{j(j+1)}$ & $1$ & $N_s - 1$ & $2j+1$\\
\cline{2-5}
& $\frac{1}{3}(j+1)$ & $0$ & $N_s - 1$ & $2j+3$\\
\hline
& $\frac{1}{3} \sqrt{j^2+6j+6}$ & $1$ & $N_s - 1$ & $2j-1$\\
\cline{2-5}
$\eta_s = -1$ & $\frac{1}{3}\sqrt{j^2+j+6}$ & $1$ & $N_s - 1$ & $2j+1$\\
\cline{2-5}
& $\frac{1}{3}\sqrt{j^2-4j+1}$ & $0$ & $N_s - 1$ & $2j+3$\\
\hline
\end{tabular}
}
&
\subtable[fermions\label{table:psi:diag}]{
\begin{tabular}{|c|c|r@{$\leq j \leq$}l|c|}
\hline
& mass & \multicolumn{2}{c}{range of $j$} & degeneracy \\
\hline
$\eta_s=1$ & $\frac{1}{3}\left(j+\frac{3}{4}\right)$ & $0$ & $N_s-1$
 & $2j+2$ \\
\cline{2-5}
& $\frac{1}{3}\left(j+\frac{1}{4}\right)$ & $1$ & $N_s-1$
 & $2j$\\
\hline
$\eta_s=-1$ & $\left\lvert\frac{j}{3}-\frac{1}{4}\right\rvert$ & $0$ & $N_s-1$
 & $2j+2$ \\
\cline{2-5}
& $\frac{j}{3}+\frac{7}{12}$ & $1$ & $N_s-1$
 & $2j$ \\
\hline
\end{tabular}
} 
\\ \\ 
\subtable[$Z^0$\label{table:Z0:diag}]{
\begin{tabular}{|c|r@{$\leq j \leq$}l|c|}
\hline
mass & \multicolumn{2}{c|}{range of $j$} & degeneracy \\
\hline
$\frac{1}{3}\sqrt{j(j+1)}$ & $0$ & $N_s-1$
 & $2j+1$ \\
\hline
\end{tabular}
} 
&
\subtable[$Z^a$\label{table:Za:diag}]{
\begin{tabular}{|c|r@{$\leq j \leq$}l|c|}
\hline
mass & \multicolumn{2}{c|}{range of $j$} & degeneracy \\
\hline
$\frac{1}{3}\left(j+\frac{1}{2}\right)$ & $0$ & $N_s-1$
 & $2j+1$ \\
\hline
\end{tabular}
} 
\\ \\ \end{tabular}
\begin{tabular}{c} 
\subtable[ghosts\label{table:ghosts:diag}]{
\begin{tabular}{|c|r@{$\leq j \leq$}l|c|}
\hline
mass & \multicolumn{2}{c|}{range of $j$} & degeneracy \\
\hline
$\frac{1}{3}\sqrt{j(j+1)}$ & $0$ & $N_s-1$
 & $2j+1$ \\
\hline
\end{tabular}
} 
\end{tabular}
\caption{The mass spectrum of the diagonal blocks.
\label{table:diag}}
\end{center}
\end{table}

As for the ghost part of the action, there is no difference between
$s=1$ and $s=2$, and for both $\eta_s=\pm 1$ the masses of the
ghosts are $\frac{1}{3}\sqrt{j(j+1)}$ with $j=0,\dots,N_s-1$ and
$m=-j,\dots,j$. This completes our presentation of the mass spectrum
of the diagonal fluctuations;
see Table~\ref{table:diag}. Note that this spectrum is independent
of the antimembrane's motion $x^a(t)$ in the $x^4,\dots,x^9$
directions. As a result, $x^a(t)$ (and its time-derivative) does not
appear in the part of the one-loop effective potential coming from
the diagonal fluctuations, which we turn to below.

\subsubsection{One-Loop Effective Potential For a Single
Antimembrane}\label{subsubsection: self one-loop}

Using the mass spectrum of the diagonal fluctuations found in
Section \ref{subsubsection diagonal} we can now compute the one-loop
potential for the membrane alone and also for the antimembrane alone---%
\ie\ that part of the potential that does not include
the interaction between the two.
For the membrane, the mass spectrum we gave above is the
same as that in \cite{Shin:2003np}, for which
$V_\text{eff}^\text{one-loop}$ vanishes as expected (since it is
supersymmetric). On the other hand, for the antimembrane all the
supersymmetries are broken and the one-loop potential does not
vanish. Before writing down the formula for the effective potential,
we first undo the rescaling done on the fields and time in eqn.
(\ref{eqn: rescaling}) by replacing $v\to\frac{v}{\mu}, V_\text{eff}\to
\mu V_\text{eff}$, and also restore powers of $\alpha$. One then finds
\begin{equation} \label{eqn:mbar Veff}
\begin{split}
\bar{V}_\text{eff}^\text{one-loop}=-\frac{1}{2}\mu\Bigg[ &
\sum_{j=0}^{N_2-1}(2j+1)\frac{1}{3}\sqrt{j(j+1)}
+6\sum_{j=0}^{N_2-1}(2j+1)\frac{1}{3}\left(j+\frac{1}{2}\right)\\
&
+\sum_{j=1}^{N_2-1}(2j-1)\frac{1}{3}\sqrt{j^2+6j+6}+\sum_{j=1}^{N_2-1}(2j+1)
\frac{1}{3}\sqrt{j^2+j+6}\\
&+\sum_{j=0}^{N_2-1}(2j+3)\frac{1}{3}\sqrt{j^2-4j+1}
-4\sum_{j=0}^{N_2-1}(2j+2)\left\lvert\frac{j}{3}-\frac{1}{4}\right\rvert\\
&-4\sum_{j=1}^{N_2-1}2j\left(\frac{j}{3}+\frac{7}{12}\right)
-2\sum_{j=0}^{N_2-1}(2j+1)\frac{1}{3}\sqrt{j(j+1)} \Bigg].
\end{split}
\end{equation}

To expand the above summation (\ref{eqn:mbar Veff}) for large $N_2$,
we use the Euler-Maclaurin formula,
\BE\label{eqn: Euler-Maclaurin}
\sum_{k=1}^{n-1} f(k)&= &\int_0^n
f(k)dk-\frac{1}{2}\left[f(n)+f(0)\right]
+\sum_{m=1}^\infty\frac{B_{2m}}{(2m)!}
\left[f^{(2m-1)}(n)-f^{(2m-1)}(0)\right],
\EE
where $f^{(m)}(k)$
stands for the $m$th-derivative of $f$, and $B_{2m}$ is the
Bernoulli number ($\frac{B_2}{2!}=\frac{1}{12}$,
$\frac{B_4}{4!}=-\frac{1}{720}$, $\frac{B_6}{6!}=\frac{1}{30240}$,
$\frac{B_8}{8!}=-\frac{1}{1209600}$, and so on). Applying this to
the summation (\ref{eqn:mbar Veff}) (using
{\sf Mathematica}$^\circledR$), we find
that all terms with positive powers of $N_2$ cancel, leaving
\BE\label{eqn: barVeff}
\bar{V}_\text{eff}^\text{one-loop}(r'_0)
=-i\left[\frac{7\mu}{6}(2\sqrt{2}+\sqrt{3})\right]
+\left\{-\frac{\mu}{2}\left[b+\frac{87}{4}\frac{1}{N_2}
   +\order{\frac{1}{N_2^2}}\right]\right\}.
\EE
Note that $\bar{V}_\text{eff}^\text{one-loop}$ has a constant imaginary part
coming from the twenty-one tachyon modes in the antimembrane's
fluctuations along the $X^{1,2,3}$ directions, which gives a
constant decay rate. The pure number $b$ in the real part of
$\bar{V}_\text{eff}^\text{one-loop}$ has an
approximate value of $-4.56$ and is
just the zero-point energy of the nonsupersymmetric antimembrane.

\subsubsection{Off-Diagonal Fluctuations}\label{subsubsection off-diag}
For any $N_r\times N_s$ matrix $M$ define \BE \anti{J^i}{M} \equiv
J^i_{(r)}M+MJ^i_{(s)}, \EE and \BE \com{J^i}{M} \equiv
J^i_{(r)}M-MJ^i_{(s)}. \EE (When $N_r=N_s$, this notation goes
without saying, of course.) One useful identity is \BE
\anti{J^i}{\anti{J^i}{M}}=2(\Lambda_{(r)}+\Lambda_{(s)})M -
\com{J^i}{\com{J^i}{M}}. \EE with $\Lambda_{(r)}=\frac{N_r^2-1}{4}$
being the Casimir $J^i_{(r)}J^i_{(r)}=\Lambda_{(r)}\cdot
I_{N_r\times N_r}$.

The off-diagonal fluctuations give the interaction between the
membrane and the antimembrane. Let us first look at the bosonic
part. As can be readily seen, the $\Phi^0,\Phi^a$ part of the action
and the $\Phi^{1,2,3}$ part are not coupled, \ie,
$(S_\text{boson})_\text{o.d.}=(S_\text{boson})_{\text{o.d.},
0a}+(S_\text{boson})_{\text{o.d.},123}$, with
\begin{multline} \label{eqn: SB0a}
(S_\text{boson})_{\text{o.d.},0a}
=\Tr\Bigg(
\dot{\Phi}^a\dot{\Phi}^{a\dagger}
-2i\frac{v^a}{\mu}(\Phi^{0\dagger}\Phi^a
-\Phi^{a\dagger}\Phi^0)
-\left(\frac{1}{3}\right)^2 \com{J^i}{\Phi^a}\com{J^i}{\Phi^{a\dagger}} \\*
-\dot{\Phi}^0\dot{\Phi}^{0\dagger}
+\left(\frac{1}{3}\right)^2 \com{J^i}{\Phi^0}\com{J^i}{\Phi^{0\dagger}}
+\left[2\left(\frac{1}{3}\right)^2(\Lambda_{1}+\Lambda_{(2)})+
           \frac{x^bx^b}{\mu^2}\right]\Phi^0\Phi^{0\dagger} \\*
-\left[2\left(\frac{1}{3}\right)^2(\Lambda_{1}+\Lambda_{(2)})
       +\left(\frac{1}{6}\right)^2
+\frac{x^bx^b}{\mu^2}\right]\Phi^a\Phi^{a\dagger}\Bigg),
\end{multline}
(with $v^a\equiv\dot{x}^a$) and
\begin{multline} \label{eqn: SB123}
(S_\text{boson})_{\text{o.d.},123}
=\Tr\Bigg(
\dot{\Phi}^i\dot{\Phi}^{i\dagger}
-\left(\frac{1}{3}\right)^2 \anti{J^i}{\Phi^j}\anti{J^i}{\Phi^{j\dagger}}
-\frac{x^bx^b}{\mu^2}\Phi^i\Phi^{i\dagger}\\
+2\left(\frac{1}{3}\right)^2 i\epsilon^{ijk}
     \com{J^k}{\Phi^i}\Phi^{j\dagger}
-i\epsilon^{ijk}\left(\frac{1}{3}\right)\anti{J^i}{\Phi^j}\Phi^{k\dagger}
-\left(\frac{1}{3}\right)^2\Phi^i\Phi^{i\dagger}\Bigg).
\end{multline}

The masses of $\Phi^0,\Phi^a$ can be readily found. Making use of
the $SO(6)$ symmetry to set $v^a=v\delta^{a9}$, we see that
$\Phi^0,\Phi^9$ are coupled, while $\Phi^{4,5,6,7,8}$ are decoupled.
One then expands the fields using the $N_1\times N_2$ matrix
spherical harmonics \BE \Phi^I=\sum_{j=\frac{\left\lvert
N_1-N_2\right\rvert}{2}}^{\frac{N_1+N_2}{2}-1}\sum_{m=-j}^j
\Phi^I_{jm}Y_{jm}.\EE The masses of $\Phi^{4,5,6,7,8}$ are then
immediately seen to be
\BE
\sqrt{2\left(\frac{1}{3}\right)^2(\Lambda_{(1)}+\Lambda_{(2)})
+\left(\frac{1}{6}\right)^2+\frac{x^bx^b}{\mu^2}-\left(\frac{1}{3}\right)^2j(j+1)
},
\EE with $j=\frac{\left\lvert
N_1-N_2\right\rvert}{2},\dots,\frac{N_1+N_2}{2}-1$ and $m=-j,\dots,j$.
For $\Phi^0,\Phi^9$, we first Wick rotate $A\to iA$
(which means $\Phi^0\to i\Phi^0$ and $\Phi^{0\dagger}\to i\Phi^{0\dagger}$) and
$v\to -iv$. Then one finds that the mass squared matrix for
$\Phi^0,\Phi^9$ is given by
\BE
\begin{pmatrix}\tilde{\Delta}&i\frac{2v}{\mu}\\
-i\frac{2v}{\mu}&\tilde{\Delta}+\left(\frac{1}{6}\right)^2\end{pmatrix},
\EE
where we have expanded using matrix spherical harmonics and
suppressed the $jm$ subscripts (since different $jm$ components do
not mix), and $\tilde{\Delta}\equiv
2\left(\frac{1}{3}\right)^2\left(\Lambda_{(1)}+\Lambda_{(2)}\right)+\frac{x^bx^b}{\mu^2}-\left(\frac{1}{3}\right)^2
j(j+1)$. Diagonalizing this $2\times 2$ matrix we get the mass of
$\Phi^0,\Phi^9$ \BE
\left[\tilde{\Delta}+\frac{1}{72}\pm\frac{1}{72}\sqrt{1+\left(\frac{144v}{\mu}\right)^2}\right]^{1/2},\EE
with the range of $(jm)$ being $j=\frac{\left\lvert
N_1-N_2\right\rvert}{2},\dots,\frac{N_1+N_2}{2}-1$ and $m=-j,\dots,j$.

Finding the masses of $\Phi^{1,2,3}$ involves substantial work and
the details are given in Appendix~\ref{sec:Phimass}.
Those masses contain $x^a(t)$
but not its time-derivative $v$.

The mass spectrum of the off-diagonal fermionic fluctuation $\chi$
is computed in Appendix~\ref{sec:chimass} and summarized in
eqns.~(\ref{eqn: fermion mass spectrum}) and (\ref{eqn: fermion mass spectrum
detail}).

The off-diagonal ghost part of the action is given by
\begin{multline}
(S_\text{ghost})_\text{o.d.}
=\Tr\Bigg((\dot{\bar{C}}\dot{C}^\dagger+\dot{\bar{C}}^\dagger\dot{C})
-\left(\frac{1}{3}\right)^2\anti{J^i}{\bar{C}}\anti{J^i}{C^\dagger}
-\frac{x^ax^a}{\mu^2}\bar{C}C^\dagger \\
-\left(\frac{1}{3}\right)^2\anti{J^i}{\bar{C}^\dagger}\anti{J^i}{C}
-\frac{x^ax^a}{\mu^2}\bar{C}^\dagger C\Bigg).
\end{multline}
Using
\begin{equation}
\begin{split}
\Tr\left(-\anti{J^i}{\bar{C}}\anti{J^i}{C^\dagger}\right)
&=\Tr\left(C^\dagger \anti{J^i}{\anti{J^i}{\bar{C}}}
\right)\\
&=\Tr\left(-
2(\Lambda_{(1)}+\Lambda_{(2)})\bar{C}C^\dagger -
\com{J^i}{\bar{C}}\com{J^i}{C^\dagger}\right),
\end{split}
\end{equation}
one readily finds that the masses of $\bar{C}$ and $C^\dagger$ (with
their complex conjugates being $\bar{C}^\dagger$ and $C$ respectively) are
the same and are given by \BE
\sqrt{2\left(\frac{1}{3}\right)^2(\Lambda_{(1)}+\Lambda_{(2)})+\frac{x^ax^a}{\mu^2}
-\left(\frac{1}{3}\right)^2 j(j+1)},\EE with $j=\frac{\left\lvert
N_1-N_2\right\rvert}{2},\dots,\frac{N_1+N_2}{2}-1$ and $m=-j,\dots,j$.

\subsubsection{Interaction Between the Membrane and the
Antimembrane}\label{subsubsection interaction}

The part of the one-loop potential describing the interaction
between the membrane block and the antimembrane block is obtained
using the mass spectrum of the off-diagonal fluctuations worked out
in section \ref{subsubsection off-diag}. This is done by writing
down the sum-over-mass expression (\ref{eqn: sum over mass}), using
the Euler-Maclaurin formula (\ref{eqn: Euler-Maclaurin}), replacing
$N_2$ by $\frac{6r_0'}{\alpha\mu}$, $N_1$ by
$\frac{6r_0}{\alpha\mu}$, then expanding in powers of $\alpha$ and
in the end dropping terms at quadratic or higher orders in $\alpha$.
The reason for this $\alpha$-expansion is the same as explained in
\cite{Lee:2004kv}: as we shall soon see in Section \ref{section:
sugra Llc}, the interaction on the supergravity side is
$\order{\alpha^1}$; since here we are only interested in comparing matrix
theory predictions to supergravity results, on the matrix theory
side we can also just keep order $\alpha^1$. Indeed, higher
orders of $\alpha$ come with higher powers of $\frac{1}{r}$, thus
becoming important only at short distances; these are matrix theory
corrections beyond supergravity.%
\footnote{In Section
\ref{subsubsection: self one-loop}, we did a $1/N_2$ expansion
to compute the effective
potential for a single antimembrane.
Since the only $\alpha$-dependence appears via
$N_2=\frac{6r_0'}{\alpha\mu}$, this amounted to an
$\alpha$-expansion. Of course it has to be this way; despite the
fact that diagonal fluctuations were considered there and
off-diagonal fluctuations here, they both belong to the same
quantity, namely the one-loop effective potential.}

We shall compute the velocity-dependent part of the effective
action, $V_\text{eff,$v$-dep.}^\text{one-loop}$, which can be seen
to receive contributions only from the masses of $\Phi^0,\Phi^9$ and the
fermion $\chi$. After restoring powers of $\alpha$ and $\mu$ as
described above eqn.~(\ref{eqn:mbar Veff}), the sum-over-mass for
$\Phi^0,\Phi^9$, $\chi$ is (assuming that $N_2>N_1$, \ie\ that the
difference between the radii of the antimembrane and the membrane
$w\equiv r_0'-r_0>0$, and defining $z\equiv\sqrt{x^ax^a}$ to be
the separation of the two objects in the $x^4,\dots,x^9$ directions)
\begin{equation}
\begin{split}
V_\text{eff,$v$-dep.}^\text{one-loop}=-\frac{1}{2\alpha}\Bigg\{&
2\sum_{j=\frac{\left\lvert
N_1-N_2\right\rvert}{2}}^{\frac{N_1+N_2}{2}-1}(2j+1)\Bigg[\sqrt{z^2+\left(\frac{\alpha\mu}{3}\right)^2
\left[\frac{N_1^2+N_2^2-2}{2}-j(j+1)\right]+\frac{(\alpha\mu)^2}{72}+\frac{1}{72}\sqrt{(\alpha\mu)^4+
\left(144\alpha v\right)^2}}\\
& \qquad \qquad 
+\sqrt{z^2+\left(\frac{\alpha\mu}{3}\right)^2
\left[\frac{N_1^2+N_2^2-2}{2}-j(j+1)\right]+\frac{(\alpha\mu)^2}{72}-\frac{1}{72}\sqrt{(\alpha\mu)^4+
\left(144\alpha v\right)^2}}\ \Bigg] \\
&
-4(N_1+N_2)\sqrt{z^2+(\alpha\mu)^2\left(\frac{N_2-N_1}{6}-\frac{1}{4}\right)^2+\alpha
v}\\
&-4(N_2-N_1)\sqrt{z^2+(\alpha\mu)^2\left(-\frac{N_1+N_2}{6}-\frac{1}{4}\right)^2+\alpha
v}\\
&-4\sum_{l=\frac{\left\lvert N_1-N_2
\right\rvert}{2}}^{\frac{N_1+N_2}{2}-2}2(l+1)\Bigg[\sqrt{z^2+(\alpha\mu)^2\left(\frac{1}{3}
\sqrt{\frac{N_1^2+N_2^2}{2}-(l+1)^2}-\frac{1}{4} \right)^2+\alpha
v}\\
& \qquad \qquad 
+\sqrt{z^2+(\alpha\mu)^2\left(-\frac{1}{3}
\sqrt{\frac{N_1^2+N_2^2}{2}-(l+1)^2}-\frac{1}{4} \right)^2+\alpha
v}\Bigg]\\
 &
-4(N_1+N_2)\sqrt{z^2+(\alpha\mu)^2\left(\frac{N_2-N_1}{6}-\frac{1}{4}\right)^2-\alpha
v}\\
&-4(N_2-N_1)\sqrt{z^2+(\alpha\mu)^2\left(-\frac{N_1+N_2}{6}-\frac{1}{4}\right)^2-\alpha
v}\\
&-4\sum_{l=\frac{\left\lvert N_1-N_2
\right\rvert}{2}}^{\frac{N_1+N_2}{2}-2}2(l+1)\Bigg[\sqrt{z^2+(\alpha\mu)^2\left(\frac{1}{3}
\sqrt{\frac{N_1^2+N_2^2}{2}-(l+1)^2}-\frac{1}{4} \right)^2-\alpha
v}\\
& \qquad \qquad 
+\sqrt{z^2+(\alpha\mu)^2\left(-\frac{1}{3}
\sqrt{\frac{N_1^2+N_2^2}{2}-(l+1)^2}-\frac{1}{4} \right)^2-\alpha
v}\Bigg]\Bigg\}.
\end{split}
\end{equation}
Carrying out the procedures outlined above, we
find that, for the velocity-dependent part of the above
$V_\text{eff,$v$-dep.}^\text{one-loop}$,  terms at $O(\alpha^0)$ as
well as terms with negative powers of $\alpha$ all cancel out, and
and the final result is (after Wick rotating back, \ie\ $v\to iv$)
\BE\label{eqn: Vmatrix v-dep}
V_\text{eff,$v$-dep.}^\text{one-loop}=\frac{9\alpha}{\mu^2(w^2+z^2)^{5/2}}\left\{
\frac{3}{8}v^4+v^2r_0^2\mu^2\left[\frac{1}{3}+\frac{5}{6}\frac{w}{r_0}
+\frac{1}{48}\frac{26w^2+z^2}{r_0^2}\right]\right\}.\EE
The final step in getting the result (\ref{eqn: Vmatrix v-dep}) is to do
an additional expansion called the ``near-membrane expansion'' in which we expand 
in the parameter $\frac{\xi}{r_0}$ (where $\xi\equiv\sqrt{w^2+z^2}$) and then only keep the leading terms (which are the terms 
that diverge as $\xi\to 0$). The only reason for doing this near-membrane expansion is that
on the supergravity side we have only worked to leading orders of this expansion (see Section \ref{section: sugra
Llc} below and \cite{Lee:2004kv}) and
we want to compare the result here with the result there. Without doing this expansion, we
would obtain the analog of the interpolating potential given in Section 6 of \cite{Lee:2004kv}, 
which would be valid regardless of whether the separation $\xi$ between the membrane and 
the anti-membrane is much smaller than their radii or not. Additionally, we have set $v$ to be of order
$\mu\xi$, in accordance with Section \ref{section: sugra
Llc} below.

\section{Interaction Calculated on the Supergravity Side}\label{section: sugra
Llc}

On the linearized supergravity side we use the source-probe
analysis, treating the membrane as the source and the antimembrane
as the probe%
\footnote{Since the antimembrane is treated as a probe
it does not contribute to the stress-energy tensor, and thus
integrability of Einstein equation does not require its trajectory
to satisfy equation of motion. See also
\cite{Lee:2003kf,Lee:2004kv}.}
. The metric and gauge field
perturbations $h_{\mu\nu},a_{\mu\nu\rho}$ produced by the source
membrane were computed in \cite{Lee:2004kv}. As pointed out in
Section \ref{section: intro antimembrane}, to get the lightcone
Lagrangian $\bar{L}_\text{l.c.}$ for the antimembrane we just have
to take the membrane $L_\text{l.c.}$ given in \cite{Lee:2004kv} and
flip the sign of the terms containing $a_{\mu\nu\rho}$. We shall
make the assumption that $v$ is of order $\mu\xi, \xi\equiv \sqrt{w^2+z^2}$.
This is because
on-shell trajectories of the antimembrane are in general elliptical
orbits in the $x^{4,...,9}$ directions and have $v\sim \mu z\sim
\mu\xi$. Although here we do not require the trajectory to be
on-shell---the orbit can be of arbitrary shape---we
choose it not to be ``too off-shell'' by requiring its velocity to
be of the same order of magnitude as those of on-shell trajectories.
Finally, as in \cite{Lee:2004kv}, since we are only interested in
the part of the membrane/antimembrane interaction that diverges as
the two objects get closer and closer, in the expression for
$\bar{L}_\text{l.c.}$ we shall only keep the terms that are singular
as $\xi\to 0$ which are the leading terms in the near-membrane expansion.

The result, upon writing $\bar{L}_\text{l.c.}$ as the sum of a
velocity-dependent part $\bar{L}_\text{l.c., $v$-dep.}$ and a
velocity-independent part $\bar{L}_{\text{l.c., $v$-indep.}}$, is
\begin{align}
\bar{L}_\text{l.c., $v$-dep.}&=\left(\int d\theta d\phi
\Pi_-\Delta\right)\frac{1}{(w^2+z^2)^{5/2}}\left\{
\frac{3}{8}v^4+v^2r_0^2\mu^2\left[\frac{1}{3}+\frac{1}{2}\frac{w}{r_0}
+\frac{1}{48}\frac{2w^2+z^2}{r_0^2}\right]\right\},
\\
\begin{split}
\bar{L}_\text{lc, $v$-indep.}&=\left(\int d\theta d\phi
\Pi_-\Delta\right)\frac{\mu^4r_0^4}{9(w^2+z^2)^{5/2}}\Bigg\{
\frac{2}{3}+\frac{8}{3}\frac{w}{r_0}+\frac{(32w^2-11z^2)}{12r_0^2} \\ & \qquad
+\frac{w(8w^2-5z^2)}{8r_0^3}+\frac{(128w^4+252w^2z^2+125z^4)}{384r_0^4}\Bigg\},
\end{split}
\end{align}
where the quantity $\Delta$, which is a proportionality constant
in $h_{\mu\nu},a_{\mu\nu\rho}$, is given by
$\frac{\kappa_{11}^2T}{16\pi^4R\left(\frac{\mu r_0}{3}\right)}$ (see
eqn.~(76) of \cite{Lee:2004kv}). Now recalling that
$r_0'=\frac{\mu\tilde{p}^+}{3T}=\frac{\mu\Pi_-}{3T\sin\theta}$ gives
$\Pi_-=\frac{3r_0'T\sin\theta}{\mu}=
\frac{3r_0T\sin\theta}{\mu}\left(1+\frac{w}{r_0}\right)$,
$\kappa_{11}^2=\frac{16\pi^5}{M^9}$, $T=\frac{M^3}{2\pi}$, and
$\frac{1}{M^3R}=\alpha$,  we find
\BE \int d\theta\, d\phi\,
\Pi_-\Delta=\frac{9\alpha}{\mu^2}\left(1+\frac{w}{r_0}\right).\EE
Plugging in this value of $\int d\theta d\phi \Pi_-\Delta$ yields
\begin{align} \label{eqn: sugra Veff v-dep}
\bar{L}_\text{l.c.,$v$-dep.}&=\frac{9\alpha}{\mu^2(w^2+z^2)^{5/2}}\left\{
\frac{3}{8}v^4+v^2r_0^2\mu^2\left[\frac{1}{3}+\frac{5}{6}\frac{w}{r_0}+\frac{26w^2+z^2}{48r_0^2}\right]\right\},
\\ \label{eqn: sugra Veff v-indep}
\bar{L}_\text{lc,$v$-indep.}&=\frac{\alpha\mu^2r_0^4}{(w^2+z^2)^{5/2}}\Bigg\{
\frac{2}{3}+\frac{10}{3}\frac{w}{r_0}+\frac{(64w^2-11z^2)}{12r_0^2}
+\frac{w(88w^2-37z^2)}{24r_0^3}
+\frac{(512w^4+12w^2z^2+125z^4)}{384r_0^4}\Bigg\},
\end{align}
where in the final expressions of the $\bar{L}$'s we have again only
kept terms that are singular as $\xi\to 0$.

Comparing the matrix theory result (\ref{eqn: Vmatrix v-dep}) with
the supergravity result (\ref{eqn: sugra Veff v-dep}), we see that
they completely agree. That is to say, at leading order of large
$r_0$ (\ie\ in the flat space limit, which is given by
$r_0\to\infty$, $\mu\to 0$, holding $\mu r_0$ fixed) they both
reduce to
\BE
\frac{9\alpha}{\mu^2(w^2+z^2)^{5/2}}
\left( \frac{3}{8}v^4+\frac{1}{3}v^2r_0^2\mu^2\right).
\EE
This reproduces the flat space agreement;
furthermore, the \ppwave\ corrections to
the potentials, \ie\ the
$\left(\frac{5}{6}\frac{w}{r_0}+\frac{26w^2+z^2}{48r_0^2}\right)$
terms, also agree. Thus, the flat space agreement is extended to the \ppwave.
This is quite remarkable.

Let us also compare the membrane-antimembrane interaction found
above with interaction of other objects in the \ppwave. The
velocity-dependent part of the membrane-membrane interaction is
(eqn.~(95) of \cite{Lee:2004kv})
\BE\label{eqn: m-m potential}
L_{\text{membrane-membrane}}
=\frac{9\alpha}{\mu^2(w^2+z^2)^{5/2}}\left\{\frac{3}{8}v^4-v^2r_0^2\mu^2
\left[\frac{2w^2+5z^2}{144r_0^2}\right]\right\}.
\EE
Although in the
$X^4,\dots,X^9$ directions the membrane-antimembrane pair are point
particles like gravitons, the interaction in these directions is not
the same as that of gravitons. In fact, the ratio of the coefficient
of the $v^2\mu^2z^2$ term to that of the $v^4$ term for the
membrane-antimembrane interaction (\ref{eqn: sugra Veff v-dep}) is
$\frac{1}{18}$, while for the graviton-graviton interaction this
ratio is different, being given by $\frac{7}{90}$ (see equation (97)
of \cite{Lee:2004kv}).

\section{Discussion}

In this work, we examined a spherical membrane and antimembrane in the
M-theory \ppwave\ using both M(atrix) theory, and supergravity.  We
have seen that the one-loop potential reproduces the interactions seen
in the supergravity from a probe anlaysis.  This remarkable agreement,
for a nonsupersymmetric system, does not just provide more evidence
for matrix theory.  It also leads to additional interesting questions as,
given the lack of supersymmetry in the system, one might have expected
the potential to be renormalized towards (na\"{\i}ve) disagreement.
As mentioned in the Introduction, it is possible that the membrane-antimembrane pair in \ppwave\
is {\em approximately\/} supersymmetric in a way similar to
a membrane-antimembrane pair in flat space.
That would explain the agreement we have found.

We have put more emphasis on the interaction between
the membrane and antimembrane as a way of testing the gauge/gravity
duality in a nonsupersymmetric setting, and only briefly talked
about of the dynamics of a single antimembrane in the \ppwave. However,
it is worth pointing out the antimembrane by itself deserves further
investigation. First, we have found that its fluctuations have
tachyon modes and hence one would like to better understand what is the
final product of the corresponding decay process. One natural guess
is these tachyon modes cause the antimembrane to deform (and perhaps
finally disintegrate), during which gravitational as well as three-form
gauge field radiation is emitted.
Alternatively, one might suppose that the antimembrane collapses and passes through itself, thereby inverting its dipole to become a membrane,
\cf~\cite{Das:2005vd}, perhaps emitting radiation in
the process.
(In fact, the first guess is a special case of the second, in which the
final membrane state is in the trivial SU(2) vacuum.)
It would be interesting to show
what happens by a concrete computation.
Secondly, we have put the antimembrane off-shell at a constant
radius $r_0'=\frac{\mu\tilde{p^+}}{3T}$. By looking at the
antimembrane's Lagrangian (\ref{eqn: Stree sugra}) we see that the
trajectory solving the classical e.o.m., which depicts a collapsing
antimembrane, is given by elliptic integrals. The recent work
\cite{Papageorgakis:2005zf} considered spherical D-branes in flat
space which collapse due to its tension. The trajectory there
possesses a large-small duality relating $r$ to $1/r$, which comes
from complex multiplication properties of the Jacobi elliptic
functions. It is therefore not inconceivable that our antimembrane
will also exhibit this large-small duality, and it would be
interesting to work out the details and try to understand the
physical implications. This duality, as observed in
\cite{Papageorgakis:2005zf}, is probably a disguise of T-duality. A
better understanding of that might shed some light on the issue of
duality transformations for spherical branes discussed in the
Introduction.  This, in turn, could help explain our precise
agreement between the matrix theory and the supergravity.

We have left the matrix theory computation of the
velocity-independent part of the one-loop interaction between the
membrane and the antimembrane to future work (the interaction on the
supergravity side is given in eqn.~(\ref{eqn: sugra Veff v-indep})).
The necessary ingredients for that computation (\ie, the masses of
$\Phi^{1,2,3}$) are worked out in Appendix~\ref{sec:Phimass}, and
are fairly complicated expressions.

\acknowledgments
We thank Ofer Aharony, Dongsu Bak, Tristan
McLoughlin, and Sanjaye Ramgoolam for discussions.  X. W. is
grateful to the Third Simons Workshop in Mathematics and Physics at
YITP for hospitality and a stimulating
environment. This work is supported in part by Department of Energy
contract \#DE-FG01-00ER45832 and the National Science Foundation
grant No.~PHY-0244811.

\appendix

\section{Finding the masses of $Z^i_{(s)}$} \label{sec:Zmass}

We would like to diagonalize the $Z^i_{(s)}$ part of the action
(\ref{eqn: SBd}). This is easily done using the vector spherical
harmonics $Y^i_{jlm}$~\cite{Das:2003yq} (see also~\cite{dsv1}), for which
\begin{equation} \label{eqn: deg 1}
\begin{aligned}
\epsilon_{ijk} \com{J^j}{Y^k_{j-1,jm}} &= i (j+1) Y^i_{j-1,jm}, &
j=1,\dots,N_s-1;\;&  m=-j+1,\dots,j-1, \\
\epsilon_{ijk} \com{J^j}{Y^k_{jjm}} &= i Y^i_{jjm}, &
j=1,\dots,N_s-1;\;& m=-j,\dots,j, \\
\epsilon_{ijk} \com{J^j}{Y^k_{j+1,jm}} &= -i j Y^i_{j+1,jm}, &
j=0,\dots,N_s-1;\;&  m=-j-1,\dots,j+1.
\end{aligned}
\end{equation}
and
\begin{equation}
\com{J^i}{Y^i_{jlm}} = \sqrt{j(j+1)} \delta_{jl} Y_{lm}.
\end{equation}
One can check that the total number of vector spherical harmonics
is $3N_s^2$, as it should be.

Thus one can expand
\begin{equation}
Z^i = \sum_{j,l,m} Z_{jlm} Y^i_{jlm},
\qquad Z_{jlm} = (-1)^{j-l+m+1} Z^*_{jlm},
\end{equation}
to find that the $Z^i$ part of the action (\ref{eqn: SBd}) becomes
\begin{multline}\label{eqn: SBdnew}
\frac{1}{2}\Tr\left\{
\abs{\dot{Z}_{j-1,jm}}^2
+ \abs{\dot{Z}_{j,jm}}^2
+ \abs{\dot{Z}_{j+1,jm}}^2
-\left(\frac{1}{3}\right)^2 (j+1-\eta_s)^2 \abs{Z_{j-1,jm}}^2
-\left(\frac{1}{3}\right)^2 (1-\eta_s)^2 \abs{Z_{jjm}}^2
\right.\\\left.
-\left(\frac{1}{3}\right)^2 (j+\eta_s)^2 \abs{Z_{j+1,jm}}^2
- \left(\frac{1}{3}\right)^2 j(j+1) \abs{Z_{jjm}}^2
-(1-\eta_s)\left(\frac{1}{3}\right)^2 (j+1) \abs{Z_{j-1,jm}}^2
\right.\\\left.
-(1-\eta_s)\left(\frac{1}{3}\right)^2 \abs{Z_{jjm}}^2
+(1-\eta_s)\left(\frac{1}{3}\right)^2 j \abs{Z_{j-1,jm}}^2
\right\}.
\end{multline}
It is then easy to read off the masses of the eigenmodes.
For $\eta_s=1$,
\begin{equation}\label{eqn: mass 1}
\begin{aligned}
&\text{for
}Z_{j-1,j,m}, && \text{the mass is }
\frac{1}{3}j,\\
&\text{for }Z_{jjm}, && \text{the mass is
}\frac{1}{3}\sqrt{j(j+1)},\\
&\text{for }Z_{j+1,jm}, && \text{the mass is }\frac{1}{3}(j+1),
\end{aligned}
\end{equation}
with the degeneracies given in eqn.~(\ref{eqn: deg 1}).
(This spectrum agrees with that given in \cite{Shin:2003np}, of
course.)
For $\eta_s=-1$,
\begin{equation}\label{eqn: mass 2}
\begin{aligned}
&\text{for }Z_{j-1,jm}, && \text{the mass is }
\frac{1}{3}\sqrt{j^2+6j+6},\\
&\text{for }Z_{j,jm}, && \text{the mass is }
\frac{1}{3}\sqrt{j^2+j+6},\\
&\text{for }Z_{j+1,jm}, && \text{the mass is }
\frac{1}{3}\sqrt{j^2-4j+1},
\end{aligned}
\end{equation}
with the degeneracies again given in eqn.~\eqref{eqn: deg 1}.

\section{Finding the masses of $\Phi^{1,2,3}$---%
Fuzzy Spherical Harmonics for Anticommutators} \label{sec:Phimass}

The generalized Jacobi identities
\begin{equation} \label{genJacobi}
\anti{A}{\anti{B}{C}} = \anti{\anti{A}{B}}{C} - \com{B}{\com{A}{C}}
 = \com{\com{A}{B}}{C} + \anti{B}{\anti{A}{C}},
\end{equation}
will be used heavily in this appendix.

\subsection{A Comment on Fuzzy Spherical Harmonics}

Fuzzy spherical harmonics have been reviewed extensively
in~\cite{Das:2003yq}, and we will not repeat those comments here.
However, we will change notation relative to that reference, so that
now the SU(2) index is $i,j,\dots=1,2,3$.  We trust the reader will
not get too confused by the use of $j,l,m$ as both indices and
angular momentum quantum numbers.

As a consequence of identity (A.32) of \cite{Das:2003yq} (which
expands a product of two spherical harmonics as a sum of spherical
harmonics), observe that since (\cf.\ equation (A.16) of
\cite{Das:2003yq})
\begin{equation}
\begin{aligned}
J^\pm &= \mp\sqrt{\frac{N^2-1}{6}} Y_{1,\pm1}, & \qquad J^3 &=
\sqrt{\frac{N^2-1}{12}} Y_{10},
\end{aligned}
\end{equation}
the following identities hold\skipthis{, where it is understood that
$J^1$ acting on the left (right) is an $N_1\times N_1$ ($N_2 \times
N_2$) matrix}:
\begin{subequations}
\begin{multline}
J^3 Y^{(N_1,N_2)}_{\ell m} =
\frac{\sqrt{[(\ell+1)^2-m^2][(\ell+1)^2-\frac{(N_1-N_2)^2}{4}]
   [\frac{(N_1+N_2)^2}{4}-(\ell+1)^2]}}{2 (\ell+1) \sqrt{(2\ell+1)(2\ell+3)}}
Y_{\ell+1,m} + \frac{m [\ell(\ell+1) +
\frac{N_1^2-N_2^2}{4}]}{2\ell(\ell+1)} Y_{\ell m}
\\
+ \frac{\sqrt{(\ell^2-m^2) [\ell^2-\frac{(N_1-N_2)^2}{4}]
   [\frac{(N_1+N_2)^2}{4}-\ell^2]}}{2 \ell \sqrt{(2\ell+1)(2\ell-1)}}
Y_{\ell-1,m},
\end{multline}
\begin{multline}
J^+ Y^{(N_1,N_2)}_{\ell m} =
-\frac{\sqrt{(\ell+m+1)(\ell+m+2)[(\ell+1)^2-\frac{(N_1-N_2)^2}{4}]
   [\frac{(N_1+N_2)^2}{4}-(\ell+1)^2]}}{2 (\ell+1) \sqrt{(2\ell+1)(2\ell+3)}}
Y_{\ell+1,m+1}
\\
+ \frac{\sqrt{(\ell-m)(\ell+m+1)}
     [\ell(\ell+1) + \frac{N_1^2-N_2^2}{4}]}{2\ell(\ell+1)} Y_{\ell, m+1}
+ \frac{\sqrt{(\ell-m-1)(\ell-m) [\ell^2-\frac{(N_1-N_2)^2}{4}]
   [\frac{(N_1+N_2)^2}{4}-\ell^2]}}{2 \ell \sqrt{(2\ell+1)(2\ell-1)}}
Y_{\ell-1,m+1}.
\end{multline}
\end{subequations}%
This, the known commutators, and hermitian conjugation, is
sufficient to determine,
\begin{subequations} \label{antiJY}
\begin{multline}
\anti{J^3}{Y^{(N_1,N_2)}_{\ell m}} =
\frac{\sqrt{[(\ell+1)^2-m^2][(\ell+1)^2-\frac{(N_1-N_2)^2}{4}]
   [\frac{(N_1+N_2)^2}{4}-(\ell+1)^2]}}{(\ell+1) \sqrt{(2\ell+1)(2\ell+3)}}
Y_{\ell+1,m} + \frac{m [\frac{N_1^2-N_2^2}{4}]}{\ell(\ell+1)}
Y_{\ell m}
\\
+ \frac{\sqrt{(\ell^2-m^2) [\ell^2-\frac{(N_1-N_2)^2}{4}]
   [\frac{(N_1+N_2)^2}{4}-\ell^2]}}{\ell \sqrt{(2\ell+1)(2\ell-1)}}
Y_{\ell-1,m},
\end{multline}
\begin{multline}
\anti{J^\pm}{Y^{(N_1,N_2)}_{\ell m}} = \mp\frac{\sqrt{(\ell\pm
m+1)(\ell\pm m+2)[(\ell+1)^2-\frac{(N_1-N_2)^2}{4}]
   [\frac{(N_1+N_2)^2}{4}-(\ell+1)^2]}}{(\ell+1) \sqrt{(2\ell+1)(2\ell+3)}}
Y_{\ell+1,m\pm 1}
\\
+ \frac{\sqrt{(\ell\mp m)(\ell \pm m+1)}
     [\frac{N_1^2-N_2^2}{4}]}{\ell(\ell+1)} Y_{\ell, m \pm 1}
\pm \frac{\sqrt{(\ell\mp m-1)(\ell\mp m)
[\ell^2-\frac{(N_1-N_2)^2}{4}]
   [\frac{(N_1+N_2)^2}{4}-\ell^2]}}{\ell \sqrt{(2\ell+1)(2\ell-1)}}
Y_{\ell-1,m\pm 1}.
\end{multline}
\end{subequations}%
We will want these identities shortly.

\subsection{Fuzzy Spherical Vector Eigenvectors}

The action (\ref{eqn: SB123}) has equation of motion
\begin{equation}
\ddot{\Phi}^i = -\tfrac{N_1^2+N_2^2}{18} \Phi^i -
\left(\tfrac{\chi^i}{\mu}\right)^2 \Phi^i + \tfrac{1}{9}
\com{J^j}{\com{J^j}{\Phi^i}} + \tfrac{2}{9} i \epsilon^{ijk}
\com{J^j}{\Phi^k} - \tfrac{1}{3} i \epsilon^{ijk}
\anti{J^j}{\Phi^k}.
\end{equation}
So it is sufficient to consider the eigenvalue equation
\begin{equation} \label{eigeq}
\com{J^j}{\com{J^j}{\Phi^i}} + 2 i \epsilon^{ijk}
\com{J^j}{\Phi^k} - 3 i \epsilon^{ijk} \anti{J^j}{\Phi^k} = \lambda
\Phi^i.
\end{equation}
This is what we do now.

Let us attempt to analyse this by expanding $\Phi^i$ in the
(ordinary) fuzzy vector spherical harmonics.  To do this, we need to
know
\begin{equation} \label{genantiuneval}
\epsilon^{ijk} \anti{J^j}{Y^k_{j\ell m}}.
\end{equation}
We will work these out, in terms of vector spherical harmonics, in
turn.

\subsubsection{Preliminary Mathematical Results}
Before starting, we can obtain a useful fact, namely, that the
inner products of these vectors is
\begin{equation} \label{ipeJY}
\begin{gathered}
\tfrac{1}{\sqrt{N_1 N_2}} \Tr \epsilon^{ijk} \anti{J^j}{Y^k_{j\ell
m}} \epsilon^{ilm} \anti{J^l}{Y^{m\dagger}_{j'\ell'm'}} =
\left[\frac{N_1^2+N_2^2}{2} + \lambda_{j\ell}\right] \delta_{jj'}
     \delta_{\ell \ell'} \delta_{mm'}
- \tfrac{1}{\sqrt{N_1N_2}}
\Tr\anti{J^i}{Y^i_{j\ell m}} \anti{J^j}{Y^{j\dagger}_{j'\ell' m'}}, \\
\lambda_{j,\ell} = \begin{cases}
-j^2, & j = \ell+1, \\
-j(j+1), & j=\ell, \\
-j^2-j-1, & j=\ell-1.
\end{cases}
\end{gathered}
\end{equation}
This result is obtained using generalized Jacobi identities and the
properties (A.34)-(A.38) of \cite{Das:2003yq}. Also,
\begin{equation} \label{J2eJY}
\com{J^l}{\com{J^l}{\epsilon^{ijk}\anti{J^j}{Y^k_{j\ell m}}}} = 2 i
\com{J^i}{\anti{J^j}{Y^j_{jlm}}} - 2 i \sqrt{j(j+1)} \delta_{j,\ell}
\anti{J^i}{Y_{jm}} + \ell(\ell+1) \epsilon^{ijk}
\anti{J^l}{Y^m_{j\ell m}},
\end{equation}
upon using generalized Jacobi identities and (A.34)-(A.38) of
\cite{Das:2003yq}. So we see that knowing $\anti{J^i}{Y^i_{j\ell
m}}$ gives us valuable clues to learning~$\epsilon^{ijk}
\anti{J^j}{Y^k_{j\ell m}}$.

\begin{table}[tb]
\begin{center}
\begin{tabular}{|c|c|}
\hline
$j,\ell,m$ & $\anti{J^i}{Y^i_{j \ell m}}$ \\
\hline $j+1,j,m$ &
$\frac{\sqrt{[(j+1)^2-\frac{(N_1-N_2)^2}{4}][\frac{(N_1+N_2)^2}{4}-(j+1)^2]}}{
\sqrt{(j+1)(2j+3)}} Y_{j+1,m}$ \\
\hline
$j,j,m$ &  $\frac{N_1^2-N_2^2}{4\sqrt{j(j+1)}} Y_{jm}$ \\
\hline $j-1,j,m$ &
$-\frac{\sqrt{[j^2-\frac{(N_1-N_2)^2}{4}][\frac{(N_1+N_2)^2}{4}-j^2]}}{
\sqrt{j(2j-1)}} Y_{j-1,m}$ \\
\hline
\end{tabular}
\caption{Expressions for $\anti{J^i}{Y^i_{j\ell m}}$. \label{t:div}}
\end{center}
\end{table}

In fact, it is easy to evaluate
\begin{equation}
\anti{J^i}{Y^i_{jjm}} = \frac{1}{\sqrt{j(j+1)}}
\com{\boldsymbol{J}^2}{Y_{jm}} = \frac{N_1^2-N_2^2}{4\sqrt{j(j+1)}}
Y_{jm}.
\end{equation}
A more tedious calculation, using (cf. the signs of eqn.~(A.4) in
\cite{Das:2003yq}) $\boldsymbol{J} = \frac{1}{\sqrt{2}}
\boldsymbol{\hat{e}}_{-1} J^+ - \frac{1}{\sqrt{2}}
\boldsymbol{\hat{e}}_{+1} J^- + \boldsymbol{\hat{e}}_0 J^3$
and~\eqref{antiJY} is
\begin{equation}
\begin{split}
\anti{J^i}{Y^i_{j+1,j,m}} &= \left[
\tfrac{\sqrt{(j-m)(j-m+1)}}{2\sqrt{(j+1)(2j+1)}}
\anti{J^-}{Y_{j,m+1}}
-\tfrac{\sqrt{(j+m)(j+m+1)}}{2\sqrt{(j+1)(2j+1)}}
\anti{J^+}{Y_{j,m+1}} +
\tfrac{\sqrt{(j+1)^2-m^2}}{\sqrt{(j+1)(2j+1)}} \anti{J^3}{Y_{jm}}
\right]
\\
&=
\frac{\sqrt{[(j+1)^2-\frac{(N_1-N_2)^2}{4}][\frac{(N_1+N_2)^2}{4}-(j+1)^2]}}{
\sqrt{(j+1)(2j+3)}} Y_{j+1,m}.
\end{split}
\end{equation}
Similarly,
\begin{equation}
\anti{J^i}{Y^i_{j-1,j,m}} =
-\frac{\sqrt{[j^2-\frac{(N_1-N_2)^2}{4}][\frac{(N_1+N_2)^2}{4}-j^2]}}{
\sqrt{j(2j-1)}} Y_{j-1,m}.
\end{equation}
These are summarized in Table~\ref{t:div}.  The results of Table~\ref{t:div},
inserted into eqn.~\eqref{ipeJY}, now allow us to tabulate the inner products
\hbox{$\tfrac{1}{\sqrt{N_1 N_2}} \Tr \epsilon^{ijk} \anti{J^j}{Y^k_{j\ell m}}
\epsilon^{ilm}
 \anti{J^l}{Y^{m\dagger}_{j'\ell'm'}}$}.
These are given in Table~\ref{t:inner}.

The results of Table~\ref{t:div}, inserted into~\eqref{J2eJY}, yield,
\begin{multline} \label{JJeJYj+1}
\com{J^l}{\com{J^l}{\epsilon^{ijk}\anti{J^j}{Y^k_{j+1,j m}}}} = 2 i
\sqrt{\frac{j+2}{2j+3}} \sqrt{[(j+1)^2-\tfrac{(N_1-N_2)^2}{4}]
  [\tfrac{(N_1+N_2)^2}{4}-(j+1)^2]} Y^i_{j+1,j+1,m}
\\
+ j(j+1) \epsilon^{ijk} \anti{J^j}{Y^k_{j+1,j,m}},
\end{multline}
and
\begin{equation}
 \label{JJeJYj-1}
\com{J^l}{\com{J^l}{\epsilon^{ijk}\anti{J^j}{Y^k_{j-1,j m}}}} = -2 i
\sqrt{\frac{j-1}{2j+1}} \sqrt{[j^2-\tfrac{(N_1-N_2)^2}{4}]
  [\tfrac{(N_1+N_2)^2}{4}-j^2]} Y^i_{j+1,j+1,m}
+ j(j+1) \epsilon^{ijk} \anti{J^j}{Y^k_{j-1,j,m}}.
\end{equation}

\begin{table}[tb]
\begin{center}
\begin{tabular}{|r|c|c|}
\hline
& $j'+1,j' m$ & $j'j'm$ \\
\hline $j+1,j,m$ &
$\delta_{jj'}
\frac{[(j+1)^2-\frac{(N_1-N_2)^2}{4}][\frac{(N_1+N_2)^2}{4}-(j+1)^2]}{
(j+1)(2j+3)}$ & \\
\hline $jjm$ &
$-\delta_{j,j'+1}
\frac{(N_1^2-N_2^2)
\sqrt{[j^2-\frac{(N_1-N_2)^2}{4}][\frac{(N_1+N_2)^2}{4}-j^2]}}{
4j\sqrt{(j+1)(2j+1)}}$ &
$\delta_{jj'} \left(\frac{N_1^2+N_2^2}{2}-j(j+1)
- \frac{(N_1^2-N_2^2)^2}{16 j(j+1)}\right)$ \\
\hline $j-1,jm$ &
$\begin{matrix}
-\delta_{j-2,j'}
\frac{\sqrt{[(j-1)^2-\frac{(N_1-N_2)^2}{4}][\frac{(N_1+N_2)^2}{4}-(j-1)^2]}}{
\sqrt{j(j-1)(2j-1)(2j+1)}}
\\ \times \sqrt{[j^2-\frac{(N_1-N_2)^2}{4}][\frac{(N_1+N_2)^2}{4}-j^2]}
\end{matrix}$ &
$\delta_{j-1,j'}
\frac{(N_1^2-N_2^2)\sqrt{[j^2-\frac{(N_1-N_2)^2}{4}][\frac{(N_1+N_2)^2}{4}-j^2]
}}{4 j\sqrt{(j-1)(2j-1)}}$ \\
\hline
\end{tabular}
\begin{tabular}{|r|c|}
\hline
& $j'-1,j' m$ \\
\hline $j-1,j,m$ &
$\begin{matrix} \delta_{jj'} \left( \frac{N_1^2+N_2^2}{2}-j^2-j-1
\right. \\ \left. +
\frac{[(j+1)^2-\frac{(N_1-N_2)^2}{4}][\frac{(N_1+N_2)^2}{4}-(j+1)^2]}{
(j+1)(2j+3)}\right)\end{matrix}$
\\
\hline
\end{tabular}
\caption{Expressions for
$\tfrac{1}{\sqrt{N_1 N_2}} \Tr \epsilon^{ijk} \anti{J^j}{Y^k_{j\ell
m}} \epsilon^{ilm} \anti{J^l}{Y^{m\dagger}_{j'\ell'm'}}$ all vanish
unless $m=m'$. Note that the table is essentially symmetric; the
blank entries can be deduced from the rest of the table.
\label{t:inner}}
\end{center}
\end{table}

\subsubsection{The Antisymmetric Anticommutators of Vector Spherical Harmonics}

Now let us start working out the
anticommutators~\eqref{genantiuneval}. Note that
\begin{equation}
\epsilon^{ijk} \anti{J^j}{Y^k_{j j m}}
 = \frac{2}{\sqrt{j(j+1)}} i \anti{J^i}{Y_{jm}}
   - \frac{1}{\sqrt{j(j+1)}} \epsilon^{ijk} \com{J^j}{\anti{J^k}{Y_{jm}}}.
\end{equation}
Given eqn.~\eqref{eqn: deg 1}, it is therefore sufficient
to determine~$\anti{J^i}{Y_{jm}}$. Moreover, it is easy to work out
the inner product of the latter with $Y^i_{j'j'm'}$:
\begin{multline}
\frac{1}{\sqrt{N_1N_2}} \Tr Y^{i\dagger}_{j'j'm'} \anti{J^i}{J_{jm}}
= \frac{1}{\sqrt{N_1 N_2 j(j+1)}}
  \Tr \anti{J^i}{\com{J^i}{Y^\dagger_{j'm'}}} Y_{jm}
= \frac{1}{\sqrt{N_1 N_2 j(j+1)}}
  \Tr \com{\boldsymbol{J}^2}{Y^\dagger_{j'm'}} Y_{jm}
\\= \frac{N_1^2-N_2^2}{4\sqrt{j(j+1)}} \delta_{jj'} \delta_{mm'}.
\end{multline}
Comparing the spherical harmonics in~\eqref{antiJY} to those in
(A.33) of \cite{Das:2003yq}, it is straightforward to see that at
most $\anti{J^i}{Y_{jm}}$ also has pieces proportional to
$Y^i_{j+2,j+1,m}, Y^i_{j+1,j,m}, Y^i_{j,j-1,m}, Y^i_{j,j+1,m},
Y^i_{j-1,j,m}$ and $Y^i_{j-2,j-1,m}$. Matching coefficients yields
only
\begin{multline}
\anti{J^i}{Y_{jm}} = \frac{\sqrt{[j^2-\frac{(N_1-N_2)^2}{4}]
[\frac{(N_1+N_2)^2}{4}-j^2]}}{\sqrt{j(2j+1)}} Y^i_{j,j-1,m} +
\frac{N_1^2-N_2^2}{4\sqrt{j(j+1)}} Y^i_{jjm}
\\
-\frac{\sqrt{[(j+1)^2-\frac{(N_1-N_2)^2}{4}]
[\frac{(N_1+N_2)^2}{4}-(j+1)^2]}}{\sqrt{(2j+1)(j+1)}} Y^i_{j,j+1,m}.
\end{multline}
Thus,
\begin{multline} \label{ejYjj}
\epsilon^{ijk} \anti{J^j}{Y^k_{j j m}} = i
\frac{\sqrt{(j+1)[j^2-\frac{(N_1-N_2)^2}{4}]
[\frac{(N_1+N_2)^2}{4}-j^2]}}{j\sqrt{(2j+1)}} Y^i_{j,j-1,m} + i
\frac{N_1^2-N_2^2}{4j(j+1)} Y^i_{jjm}
\\*
+i \frac{\sqrt{j [(j+1)^2-\frac{(N_1-N_2)^2}{4}]
[\frac{(N_1+N_2)^2}{4}-(j+1)^2]}}{(j+1)\sqrt{(2j+1))}} Y^i_{j,j+1,m}.
\end{multline}

Now let us evaluate
\begin{equation}
\epsilon^{ijk} \anti{J^j}{Y^k_{j+1,j,m}}.
\end{equation}
Eqn.~\eqref{ejYjj} immediately allows us to evaluate, [transferring
the anticommutator to $Y^{i\dagger}_{j'j'm'}$ gives a
  minus sign from the $\epsilon$, but there is another minus sign upon
  using $\Tr A^\dagger B = (\Tr B^\dagger A)^*$]
\begin{equation}
\frac{1}{\sqrt{N_1 N_2}} \Tr Y^{i\dagger}_{j'j'm'} \epsilon^{ijk}
 \anti{J^j}{Y^k_{j+1,j,m}}
= i\delta_{j,j'-1} \delta_{mm'}
 \frac{\sqrt{(j+2)[(j+1)^2-\frac{(N_1-N_2)^2}{4}]
[\frac{(N_1+N_2)^2}{4}-(j+1)^2]}}{(j+1)\sqrt{(2j+3)}},
\end{equation}
and so, schematically,
\begin{equation}
\epsilon^{ijk} \anti{J^j}{Y^k_{j+1,j,m}} = i
\frac{\sqrt{(j+2)[(j+1)^2-\frac{(N_1-N_2)^2}{4}]
[\frac{(N_1+N_2)^2}{4}-(j+1)^2]}}{(j+1)\sqrt{(2j+3)}}
Y^i_{j+1,j+1,m} + Y^i_{j',j'\pm 1,m'}.
\end{equation}
Plugging into~\eqref{JJeJYj+1}, the term proportional to
$Y^i_{j+1,j+1,m}$ cancels, so we see that the remaining, schematic
terms, have eigenvalue $j(j+1)$ under the action of
$\com{J^i}{\com{J^i}{\cdot}}$.  This means that
\begin{equation}
\epsilon^{ijk} \anti{J^j}{Y^k_{j+1,j,m}} = i
\frac{\sqrt{(j+2)[(j+1)^2-\frac{(N_1-N_2)^2}{4}]
[\frac{(N_1+N_2)^2}{4}-(j+1)^2]}}{(j+1)\sqrt{(2j+3)}}
Y^i_{j+1,j+1,m} + i A_{jm} Y^i_{j+1,j,m} + i B_{jm} Y^i_{j-1,j,m},
\end{equation}
with $A$ and $B$ to be determined.  Similarly,~\eqref{JJeJYj-1}
and~\eqref{ejYjj} imply
\begin{equation}
\epsilon^{ijk} \anti{J^j}{Y^k_{j-1,j,m}} = i
\frac{\sqrt{(j-1)[j^2-\frac{(N_1-N_2)^2}{4}]
[\frac{(N_1+N_2)^2}{4}-j^2]}}{j\sqrt{(2j-1)}} Y^i_{j-1,j-1,m} + i
B_{jm} Y^i_{j+1,j,m} + i C_{jm} Y^i_{j-1,j,m},
\end{equation}
where the same symmetry that implied the coefficient of
$Y^i_{j-1,j-1,m}$ also implies that the value of $B$ is shared.

From~Table~\ref{t:inner}, (note that this is satisfied
by~\eqref{ejYjj}!) we learn that
\begin{gather}
\begin{split}
\frac{(j+2)[(j+1)^2-\frac{(N_1-N_2)^2}{4}]
[\frac{(N_1+N_2)^2}{4}-(j+1)^2]}{(j+1)^2(2j+3)} + \abs{A_{jm}}^2 +
\abs{B_{jm}}^2
\\
 = \frac{N_1^2+N_2^2}{2}-j^2
-\frac{[(j+1)^2-\frac{(N_1-N_2)^2}{4}][\frac{(N_1+N_2)^2}{4}-(j+1)^2]}{
(j+1)(2j+3)},
\end{split}
\\
\tfrac{(j-1)[j^2-\frac{(N_1-N_2)^2}{4}]
[\frac{(N_1+N_2)^2}{4}-j^2]}{j^2(2j-1)} + \abs{B_{jm}}^2 +
\abs{C_{jm}}^2 = \frac{N_1^2+N_2^2}{2}-j^2-j-1 -
\tfrac{[j^2-\frac{(N_1-N_2)^2}{4}][\frac{(N_1+N_2)^2}{4}-j^2]}{
j(2j+1)}, \\
A_{jm} B_{jm}^* + B_{jm} C_{jm}^* = 0, \\
\begin{split}
\tfrac{\sqrt{(j+2)[(j+1)^2-\frac{(N_1-N_2)^2}{4}]
[\frac{(N_1+N_2)^2}{4}-(j+1)^2]}}{(j+1)\sqrt{(2j+3)}} A_{jm} &+
\tfrac{[N_1^2-N_2^2]\sqrt{(j+2)[(j+1)^2-\frac{(N_1+N_2)^2}{4}]
  [\frac{(N_1+N_2)^2}{4}-(j+1)^2]}}{
  4(j+1)^2(j+2)\sqrt{2j+3}} \\
&= - \tfrac{[N_1^2-N_2^2]\sqrt{[(j+1)^2-\frac{(N_1+N_2)^2}{4}]
  [\frac{(N_1+N_2)^2}{4}-(j+1)^2]}}{
  4(j+1)\sqrt{(j+2)(2j+3)}},
\end{split} \\
\frac{\sqrt{(j-1)[j^2-\frac{(N_1-N_2)^2}{4}][\frac{(N_1+N_2)^2}{4}-j^2]}}{
  j\sqrt{2j-1}} B_{jm} = 0, \\
\begin{split}
\frac{\sqrt{(j-1)[j^2-\frac{(N_1-N_2)^2}{4}][\frac{(N_1+N_2)^2}{4}-j^2]}}{
  j\sqrt{2j-1}} C_{jm}
+ \frac{[N_1^2-N_2^2]\sqrt{(j-1)[j^2-\frac{(N_1+N_2)^2}{4}]
  [\frac{(N_1+N_2)^2}{4}-j^2]}}{
  4 j^2(j-1)\sqrt{2j-1}} & \\
= \frac{[N_1^2-N_2^2]
\sqrt{[j^2-\frac{(N_1-N_2)^2}{4}][\frac{(N_1+N_2)^2}{4}-j^2]}}{
  4j\sqrt{(j-1)(2j-1)}}.
\end{split}
\end{gather}
These are respectively the inner products of $(j+1,j,m)$ with
itself; $(j-1,j,m)$ with itself; $(j+1,j,m)$ with $(j-1,j,m)$,
$(j+1,j,m)$ with $(j+1,j+1,m)$, $(j+1,j,m)$ with $(j-1,j-1,m)$, and
$(j-1,j,m)$ with $(j-1,j-1,m)$.  In fact, we only need the last
three to find, finally, that
\begin{subequations} \label{eantiJY}
\begin{gather}
\begin{aligned} \label{eantiJY:j+1j}
\epsilon^{ijk} \anti{J^j}{Y^k_{j+1,j,m}} = i
\frac{\sqrt{(j+2)[(j+1)^2-\frac{(N_1-N_2)^2}{4}]
[\frac{(N_1+N_2)^2}{4}-(j+1)^2]}}{(j+1)\sqrt{(2j+3)}}
Y^i_{j+1,j+1,m}
- i \frac{N_1^2-N_2^2}{4(j+1)} Y^i_{j+1,j,m}, \\
\tfrac{\abs{N_1-N_2}}{2} \leq j \leq \tfrac{N_1+N_2}{2}-1, -j-1\leq
m \leq j+1,
\end{aligned} \\
\label{eantiJY:jj}
\begin{split}
\epsilon^{ijk} \anti{J^j}{Y^k_{j j m}} &= i
\frac{\sqrt{(j+1)[j^2-\frac{(N_1-N_2)^2}{4}]
[\frac{(N_1+N_2)^2}{4}-j^2]}}{j\sqrt{(2j+1)}} Y^i_{j,j-1,m} + i
\frac{N_1^2-N_2^2}{4j(j+1)} Y^i_{jjm}
\\
\qquad & +i \frac{\sqrt{j [(j+1)^2-\frac{(N_1-N_2)^2}{4}]
[\frac{(N_1+N_2)^2}{4}-(j+1)^2]}}{(j+1)\sqrt{(2j+1))}}
Y^i_{j,j+1,m}, \qquad \begin{smallmatrix}
\frac{\abs{N_1-N_2}}{2}+\delta_{N_1,N_2} \leq j \leq \frac{N_1+N_2}{2}-1, \\
-j \leq m \leq j \end{smallmatrix}
\end{split} \\
\label{eantiJY:j-1j}
\begin{aligned}
\epsilon^{ijk} \anti{J^j}{Y^k_{j-1,j,m}} = i
\frac{\sqrt{(j-1)[j^2-\frac{(N_1-N_2)^2}{4}]
[\frac{(N_1+N_2)^2}{4}-j^2]}}{j\sqrt{(2j-1)}} Y^i_{j-1,j-1,m}
+ i \frac{N_1^2-N_2^2}{4j} Y^i_{j-1,j,m}, \\
\tfrac{\abs{N_1-N_2}}{2} \leq j \leq \tfrac{N_1+N_2}{2}-1, j \geq 1,
-j+1\leq m \leq j-1.
\end{aligned}
\end{gather}
\end{subequations}%
In particular, the action of $\epsilon^{ijk}\anti{J^j}{\cdot}$
preserves the $jm$ values of $Y^k_{j\ell m}$! Also, the coefficients
of ``out-of-range'' vector spherical harmonics on the right-hand
sides of these equations (\eg\ the first term for $j=\frac{1}{2}$
in~\eqref{eantiJY:jj}) vanish.

\subsubsection{Eigenvectors of~\eqref{eigeq}}

We can now ``solve'' the eigen problem~\eqref{eigeq}. Before
presenting the gory details, let us present the solution in a
(hopefully) transparent manner.  The solutions are labelled%
\footnote{Alternatively, we could have chosen to use
  $\Yp_{j\ell m}$ as for the (ordinary) vector spherical harmonics---and the
  spinors, but this seemed slightly unnatural.
}
\begin{equation}
\Yp_{n j m},
\end{equation}
with corresponding eigenvalues
\begin{equation}
\lambda_{n j m} = (j+2)(j-1) + \tilde{\lambda}_{njm}.
\end{equation}
Generically, $n=-1,0,1$, $\frac{\abs{N_1-N_2}}{2}-1\leq j \leq
\frac{N_1+N_2}{2}$ and $-j \leq m \leq j$.  However, $n$ does not
run over all three values for all $j$ and $j$ does not reach the
lower limits for all $N_1, N_2$.  So more precisely, the allowed
values of $j$ are
\begin{equation} \label{allowedj}
\begin{aligned}
j &= 1 = \frac{N_1+N_2}{2}, && N_1=N_2=1, \\
\frac{\abs{N_1-N_2}}{2} = 0 \leq j &\leq N = \frac{N_1+N_2}{2},
    && N_1=N_2=N>1, \\
\frac{\abs{N_1-N_2}}{2} = \frac{1}{2} \leq j &\leq N \pm \frac{1}{2}
  = \frac{N_1+N_2}{2}, &&
    N \equiv N_1 = N_2 \pm 1, \\
\frac{\abs{N_1-N_2}}{2}-1 \leq j &\leq \frac{N_1+N_2}{2}, && N_1
\leq N_2-2 \text{ or } N_1 \geq N_2+2.
\end{aligned}
\end{equation}
That is, the lower limit of $j$ must (not surprisingly) be
nonnegative and (more surprisingly, but this follows from the
nonexistence of the vector spherical harmonic $\boldsymbol{Y}_{010}$
[in (A.33c) of \cite{Das:2003yq}, $\frac{N_1+N_2}{2}-2=-1$ is
invalid] for $N_1=N_2=1$) if $N_1=N_2=1$ then $j\neq 0$ as well.

\begin{table}[tb]
\begin{tabular*}{0.9\textwidth}{%
   c@{\extracolsep{\fill}}c@{\extracolsep{\fill}}c}
\begin{tabular}{||c|r@{ = }l||}
\hline \hline
\multicolumn{3}{||c||}{$N_1=N_2=1$} \\
\multicolumn{3}{||c||}{$j=1$} \\ 
\hline \hline
$j=1$ & $n$ & $-1$ \\
  \cline{2-3}
  & $\tilde{\lambda}_{-1,1,m}$ & 0 \\
\hline \hline
\end{tabular} 
&
\begin{tabular}{||c|r@{ = }l||}
\hline \hline
\multicolumn{3}{||c||}{$N_1=1, N_2=2$} \\
\multicolumn{3}{||c||}{$\frac{1}{2} \leq j \leq \frac{3}{2}$} \\
\hline \hline
$j=\frac{1}{2}$ & $n$ & $0$ \\
  \cline{2-3}
   & $\tilde{\lambda}_{0,\frac{1}{2},m}$ & $-3$ \\
\hline \hline
$j=\frac{3}{2}$ & $n$ & $-1$ \\
  \cline{2-3}
  & $\tilde{\lambda}_{-1,\frac{3}{2},m}$ & $\frac{3}{2}$ \\
\hline \hline
\end{tabular}
&
\begin{tabular}{||c|r@{ = }l||}
\hline \hline
\multicolumn{3}{||c||}{$N_1=2, N_2=1$} \\
\multicolumn{3}{||c||}{$\frac{1}{2} \leq j \leq \frac{3}{2}$} \\
\hline \hline
$j=\frac{1}{2}$ & $n$ & $0$ \\
  \cline{2-3}
   & $\tilde{\lambda}_{0,\frac{1}{2},m}$ & $3$ \\
\hline \hline
$j=\frac{3}{2}$ & $n$ & $-1$ \\
  \cline{2-3}
  & $\tilde{\lambda}_{-1,\frac{3}{2},m}$ & $-\frac{3}{2}$ \\
\hline \hline
\end{tabular}
\\
\subtable{\label{tab:vjnl:1}} & \subtable{\label{tab:vjnl:12}} &
\subtable{\label{tab:vjnl:21}}
\end{tabular*} \\ 
\begin{tabular*}{0.9\textwidth}{c@{\extracolsep{\fill}}c}
\begin{tabular}{||c|r@{ = }l||}
\hline\hline
\multicolumn{3}{||c||}{$N_1=1, N_2>2$} \\
\multicolumn{3}{||c||}{$\frac{N_2-1}{2}-1 \leq j \leq \frac{N_2+1}{2}$} \\
\hline\hline
$j=\frac{N_2-3}{2}$ & $n$ & $-1$ \\
   \cline{2-3}
  & $\tilde{\lambda}_{-1,\frac{N_2-3}{2},m}$ & $-\frac{3}{2}(N_2+1)$ \\
\hline\hline
$j=\frac{N_2-1}{2}$ & $n$ & $0$ \\
   \cline{2-3}
  & $\tilde{\lambda}_{0,\frac{N_2-1}{2},m}$ & $-3$ \\
\hline\hline
$j=\frac{N_2+1}{2}$ & $n$ & $-1$ \\
   \cline{2-3}
  & $\tilde{\lambda}_{-1,\frac{N_2+1}{2},m}$ & $\frac{3}{2}(N_2-1)$ \\
\hline\hline
\end{tabular} 
&
\begin{tabular}{||c|r@{ = }l||}
\hline\hline
\multicolumn{3}{||c||}{$N_1>2, N_1=1$} \\
\multicolumn{3}{||c||}{$\frac{N_1-1}{2}-1 \leq j \leq \frac{N_1+1}{2}$} \\
\hline\hline
$j=\frac{N_1-3}{2}$ & $n$ & $-1$ \\
   \cline{2-3}
  & $\tilde{\lambda}_{-1,\frac{N_1-3}{2},m}$ & $\frac{3}{2}(N_1+1)$ \\
\hline\hline
$j=\frac{N_1-1}{2}$ & $n$ & $0$ \\
   \cline{2-3}
  & $\tilde{\lambda}_{0,\frac{N_1-1}{2},m}$ & $3$ \\
\hline\hline
$j=\frac{N_1+1}{2}$ & $n$ & $-1$ \\
   \cline{2-3}
  & $\tilde{\lambda}_{-1,\frac{N_1+1}{2},m}$ & $-\frac{3}{2}(N_1-1)$ \\
\hline\hline
\end{tabular} \\
\subtable{\label{tab:vjnl:1N}} & \subtable{\label{tab:vjnl:N1}}
\end{tabular*} \\ 
\begin{center}
\begin{tabular}{c}
\begin{tabular}{||c|r@{ = }l||}
\hline \hline
\multicolumn{3}{||c||}{$N \equiv N_1=N_2-1>1$} \\
\multicolumn{3}{||c||}{$\frac{1}{2} \leq j \leq N+\frac{1}{2}$} \\
\hline \hline
$j = \frac{1}{2}$ & $n$ & $0,1$ \\
  \cline{2-3}
  & $\tilde{\lambda}_{0,\frac{1}{2},m}$ & $-\frac{3}{4} (2N+1)
          + \frac{3}{4}\sqrt{4N^2+4N-7}$ \\
  & $\tilde{\lambda}_{1,\frac{1}{2},m}$ & $-\frac{3}{4} (2N+1)
          - \frac{3}{4}\sqrt{4N^2+4N-7}$ \\
\hline \hline
$\frac{3}{2} \leq j \leq N-\frac{3}{2}$ & $n$ & $-1,0,1$ \\
  \cline{2-3}
  & $\tilde{\lambda}_{n j m}$ & eqn.~\eqref{lambda} \\
\hline\hline
$j = N - \frac{1}{2}$ & $n$ & $0,1$ \\
  \cline{2-3}
  & $\tilde{\lambda}_{0,N-\frac{1}{2},m}$ &
        $\frac{3}{4}+ \frac{3}{4}\sqrt{16N + 9}$ \\
  & $\tilde{\lambda}_{1,N-\frac{1}{2},m}$ &
        $\frac{3}{4}- \frac{3}{4}\sqrt{16N + 9}$ \\
\hline\hline
$j = N + \frac{1}{2}$ & $n$ & $-1$ \\
  \cline{2-3}
  & $\tilde{\lambda}_{-1,N+\frac{1}{2},m}$ & $\frac{3}{2}$ \\
\hline\hline
\end{tabular}
\\ 
\subtable{\label{tab:vjnl:N,N-1}}
\newcounter{savesubtable}
\setcounter{savesubtable}{\value{subtable}}
\end{tabular}
\end{center}
\caption{The allowed values of $j$ and $n$, and the corresponding
  eigenvalues for the various cases.\label{tab:vjnl}} It should be
  noted that although the values of the eigenvalues are always
  obtainable from eqn.~\eqref{lambda}, the corresponding values of $n$
  are not the same between the conventions here and there, except
  when $n$ is allowed to take on all three values.
  This table is continued$\dots$
\end{table}
\addtocounter{table}{-1} 
\begin{table}
\setcounter{subtable}{\value{savesubtable}} 
\begin{tabular*}{0.9\textwidth}{c@{\extracolsep{\fill}}c}
\begin{tabular}{||c|r@{ = }l||}
\hline \hline
\multicolumn{3}{||c||}{$N_1=N_2+1\equiv N+1>2$} \\
\multicolumn{3}{||c||}{$\frac{1}{2} \leq j \leq N+\frac{1}{2}$} \\
\hline \hline
$j = \frac{1}{2}$ & $n$ & $0,1$ \\
  \cline{2-3}
  & $\tilde{\lambda}_{0,\frac{1}{2},m}$ & $\frac{3}{4} (2N+1)
          + \frac{3}{4}\sqrt{4N^2+4N-7}$ \\
  & $\tilde{\lambda}_{1,\frac{1}{2},m}$ & $\frac{3}{4} (2N+1)
          - \frac{3}{4}\sqrt{4N^2+4N-7}$ \\
\hline \hline
$\frac{3}{2} \leq j \leq N-\frac{3}{2}$ & $n$ & $-1,0,1$ \\
  \cline{2-3}
  & $\tilde{\lambda}_{n j m}$ & eqn.~\eqref{lambda} \\
\hline\hline
$j = N - \frac{1}{2}$ & $n$ & $0,1$ \\
  \cline{2-3}
  & $\tilde{\lambda}_{0,N-\frac{1}{2},m}$ &
        $-\frac{3}{4}+ \frac{3}{4}\sqrt{16N + 9}$ \\
  & $\tilde{\lambda}_{1,N-\frac{1}{2},m}$ &
        $-\frac{3}{4}- \frac{3}{4}\sqrt{16N + 9}$ \\
\hline\hline
$j = N + \frac{1}{2}$ & $n$ & $-1$ \\
  \cline{2-3}
  & $\tilde{\lambda}_{-1,N+\frac{1}{2},m}$ & -$\frac{3}{2}$ \\
\hline\hline
\end{tabular}
&
\begin{tabular}{||c|r@{ = }l||}
\hline \hline
\multicolumn{3}{||c||}{$N_1=N_2=N>1$} \\
\multicolumn{3}{||c||}{$0 \leq j \leq N$} \\
\hline \hline
$j=0$ & $n$ & $-1$ \\
  \cline{2-3}
  & $\tilde{\lambda}_{-1,0,0}$ & $0$ \\
\hline \hline
$1\leq j \leq N-2$ & $n$ & $-1,0,1$ \\
   \cline{2-3}
   & $\tilde{\lambda}_{-1,j,m}$ & $0$ \\
   & $\tilde{\lambda}_{0,j,m}$ & $3 \sqrt{N^2 - j(j+1)}$ \\
   & $\tilde{\lambda}_{1,j,m}$ & $-3 \sqrt{N^2 - j(j+1)}$ \\
\hline \hline
$j = N-1$ & $n$ & $0, 1$ \\
   \cline{2-3}
  & $\tilde{\lambda}_{0,N-1,m}$ & $3 \sqrt{N}$ \\
  & $\tilde{\lambda}_{1,N-1,m}$ & $-3 \sqrt{N}$ \\
\hline \hline
$j = N$ & $n$ & $-1$ \\
  \cline{2-3}
  & $\tilde{\lambda}_{-1,N,m}$ & $0$ \\
\hline \hline
\end{tabular} 
\\ 
\subtable{\label{tab:vjnl:N-1,N}} & \subtable{\label{tab:vjnl:N}}
\end{tabular*} \\
\begin{center}
\begin{tabular}{c}
\begin{tabular}{||c|r@{ = }l||}
\hline \hline
\multicolumn{3}{||c||}{$\abs{N_1- N_2}>2; N_1,N_2>1$} \\
\multicolumn{3}{||c||}{$\frac{\abs{N_1-N_2}}{2}-1 \leq j \leq
   \frac{N_1+N_2}{2}$} \\
\hline \hline
$j=\frac{\abs{N_1-N_2}}{2}-1$ & $n$ & $-1$ \\
  \cline{2-3}
  & $\tilde{\lambda}_{-1,\frac{\abs{N_1-N_2}}{2}-1,m}$ &
    $\frac{3}{2} \sgn(N_1-N_2) (N_1+N_2)$ \\
\hline \hline
$j=\frac{\abs{N_1-N_2}}{2}$ & $n$ & $0,1$ \\
  \cline{2-3}
   & $\tilde{\lambda}_{0,\frac{\abs{N_1-N_2}}{2},m}$ &
     $\frac{3}{4} \sgn(N_1-N_2)(N_1+N_2)
        + \frac{3}{4}\sqrt{(N_1+N_2)^2 - 8\abs{N_1-N_2}}$ \\
   & $\tilde{\lambda}_{1,\frac{\abs{N_1-N_2}}{2},m}$ &
     $\frac{3}{4} \sgn(N_1-N_2)(N_1+N_2)
        - \frac{3}{4}\sqrt{(N_1+N_2)^2 - 8\abs{N_1-N_2}}$ \\
\hline\hline
$\frac{\abs{N_1-N_2}}{2}+1 \leq j \leq \frac{N_1+N_2}{2}-2$ & $n$ & $-1,0,1$ \\
   \cline{2-3}
   & $\tilde{\lambda}_{n,j,m}$ & eqn.~\eqref{lambda} \\
\hline \hline
$j = \frac{N_1+N_2}{2}-1$ & $n$ & $0, 1$ \\
   \cline{2-3}
  & $\tilde{\lambda}_{0,\frac{N_1+N_2}{2}-1,m}$ & $-\frac{3}{4}(N_1-N_2)
         + \frac{3}{4} \sqrt{(N_1-N_2)^2+8(N_1+N_2)}$ \\
  & $\tilde{\lambda}_{0,\frac{N_1+N_2}{2}-1,m}$ & $-\frac{3}{4}(N_1-N_2)
         - \frac{3}{4} \sqrt{(N_1-N_2)^2+8(N_1+N_2)}$ \\
\hline \hline
$j = \frac{N_1+N_2}{2}$ & $n$ & $-1$ \\
  \cline{2-3}
  & $\tilde{\lambda}_{-1,\frac{N_1+N_2}{2},m}$ & $-\frac{3}{2} (N_1-N_2)$ \\
\hline \hline
\end{tabular} 
\\
\subtable{\label{tab:vjnl:gen}}
\end{tabular}
\end{center}
\caption{\dots The rest of the table.}
\end{table}

The range of $n$ depends on $j$; as the range of $j$ depends on
$N_1$ and $N_2$, it might be most transparent to separate out those
cases at the risk of redundancy.  See also Table~\ref{tab:vjnl}.
Explicitly,
\begin{equation}
\frac{\abs{N_1-N_2}}{2} \leq j \leq \frac{N_1+N_2}{2}, m=-j\dots j,
n = \begin{cases} -1,0,1, &
   \frac{\abs{N_1-N_2}}{2}+1 \leq j \leq \frac{N_1+N_2}{2}-2, \\
-1, & j = \frac{N_1+N_2}{2}, \\
0, 1 & j = \frac{N_1+N_2}{2}-1, \\
0, 1 & j=\frac{\abs{N_1-N_2}}{2} > 0 \text{ and } N_1,N_2 > 1, \\
0 & 0 < j=\frac{\abs{N_1-N_2}}{2} \text{ and } (N_1=1 \text{ or } N_2= 1),\\
-1, & j = 0, (\text{{\em i.e.\/} $j=\frac{\abs{N_1-N_2}}{2}$ and $N_1=N_2$}),\\
-1, & j=\frac{\abs{N_1-N_2}}{2}-1 \geq 0. \\
\end{cases}
\end{equation}
That is, generically, $n=0,\pm 1$, but not all values of $j$ allow
for all three eigenvectors.  This is particularly complicated as one
must watch for possible overlap of the various na\"{\i}vely
different cases; this is why there is a difference, for example,
between having $N_1=1$ or $N_2=1$ and having $N_1,N_2>1$.

The $\Yp_{n j m}$ are constructed to be (normalized) eigensolutions
to~\eqref{eigeq}, with eigenvalue $\lambda_{n j m}$, given by
\begin{equation}
\lambda_{n j m} = (j+2)(j-1) + \tilde{\lambda}_{n j m},
\end{equation}
where,
\begin{equation}
\tilde{\lambda}_{n j m} =
\begin{cases}
\sqrt{6} \sqrt{N_1^2+N_2^2-2 j(j+1)}
 \cos \left[\frac{\pi (2 n+1)}{3}
    - \frac{1}{3} \cos^{-1} \sqrt{\frac{27}{8}} \frac{N_1^2-N_2^2}{[
          N_1^2+N_2^2-2j(j+1)]^{3/2}} \right],
   &  \begin{matrix} n = 0,\pm 1; \\
    \frac{\abs{N_1-N_2}}{2}+1 \leq j \leq \frac{N_1+N_2}{2}-2,
    \end{matrix}  \\
-\frac{3}{2} (N_1-N_2), & n = -1; j = \frac{N_1+N_2}{2}. \\
\end{cases}
\end{equation}

Given that the commutators preserve the vector spherical harmonics,
and $\epsilon^{abc}\anti{J^b}{\cdot}$ preserves their first and last
index, we can take
\begin{equation}
\Phi^a = \left.\begin{cases}
 \alpha_{j,j-1} Y^a_{j,j-1,m} + \alpha_{j,j} Y^a_{jjm}
 + \alpha_{j,j+1} Y^a_{j,j+1,m}, &
\tfrac{\abs{N_1-N_2}}{2}+1\leq j \leq \tfrac{N_1+N_2}{2}-2,\\
\alpha_{j,j} Y^a_{jjm}  + \alpha_{j,j+1} Y^a_{j,j+1,m}, &
    N_1\neq N_2, \min(N_1, N_2) \neq 1, j=\tfrac{\abs{N_1-N_2}}{2}, \\
Y^a_{\frac{\abs{N_1-N_2}}{2} \frac{\abs{N_1-N_2}}{2} m}, &
    \min(N_1, N_2) = 1, N_1\neq N_2, j=\frac{\abs{N_1-N_2}}{2} \\
Y^a_{j,j+1,m}, & j=\tfrac{\abs{N_1-N_2}}{2}-1 \geq 0, \\
Y^a_{0 1 0}, & N_1=N_2, j=0, \\
 \alpha_{j,j-1} Y^a_{j,j-1,m} + \alpha_{j,j} Y^a_{jjm}, &
   j = \tfrac{N_1+N_2}{2}-1, \min(N_1,N_2) > 1, \\
Y^a_{j,j-1,m}, &
   j = \tfrac{N_1+N_2}{2},
\end{cases} \right\}
\qquad -j \leq m \leq j.
\end{equation}
(This indeed gives $3 N_1 N_2$ possibilities.  The constraint on the
second line, and its complement on the third line, comes from the
upper bound on $j$ in eqn.~\eqref{eantiJY:j-1j}.  There is a similar
constraint on the second-last line, with no additional complement;
indeed, if $\min(N_1,N_2)=1$ then $\frac{N_1+N_2}{2}-1 =
\frac{\abs{N_1-N_2}}{2}$, and so not only would the second-last line
be redundant with the third line, but eqn.~\eqref{eantiJY:j+1j} shows
that there is no $Y^i_{j,j-1,m}$ for this value of $j$.) Then
(A.35), (A.37) of \cite{Das:2003yq} and~\eqref{eantiJY} transform
the eigenvalue problem~\eqref{eigeq} into the eigenvalue problems,
\begin{subequations}
\begin{gather}
\begin{aligned}
\begin{pmatrix}
(j+2)(j-1)-\frac{3}{4} \frac{N_1^2-N_2^2}{j} & 3
\frac{\sqrt{(j+1)[j^2-\frac{(N_1-N_2)^2}{4}]
[\frac{(N_1+N_2)^2}{4}-j^2]}}{j\sqrt{(2j+1)}} & 0 \\
3 \frac{\sqrt{(j+1)[j^2-\frac{(N_1-N_2)^2}{4}]
[\frac{(N_1+N_2)^2}{4}-j^2]}}{j\sqrt{(2j+1)}} &
(j+2)(j-1)+\frac{3}{4} \frac{N_1^2-N_2^2}{j(j+1)} & 3
\frac{\sqrt{j[(j+1)^2-\frac{(N_1-N_2)^2}{4}]
[\frac{(N_1+N_2)^2}{4}-(j+1)^2]}}{(j+1)\sqrt{(2j+1)}} \\
0 & 3 \frac{\sqrt{j[(j+1)^2-\frac{(N_1-N_2)^2}{4}]
[\frac{(N_1+N_2)^2}{4}-(j+1)^2]}}{(j+1)\sqrt{(2j+1)}} &
(j+2)(j-1)+\frac{3}{4} \frac{N_1^2-N_2^2}{j+1}
\end{pmatrix}\\ \times
\begin{pmatrix} \alpha_{j,j-1} \\ \alpha_{j,j} \\ \alpha_{j,j+1} \end{pmatrix}
= \lambda_j
\begin{pmatrix} \alpha_{j,j-1} \\ \alpha_{j,j} \\ \alpha_{j,j+1} \end{pmatrix},
\tfrac{\abs{N_1-N_2}}{2}+1\leq j \leq \tfrac{N_1+N_2}{2}-2.
\end{aligned} \\
\begin{aligned}
\begin{pmatrix}
\frac{(N_1-N_2)^2}{4} + \frac{\abs{N_1-N_2}}{2} - 2 + 3
\sgn(N_1-N_2)\frac{N_1+N_2}{\abs{N_1-N_2}+2} & 3
\frac{\sqrt{2\abs{N_1-N_2}\left[N_1 N_2 - \abs{N_1-N_2} -1\right]}}{
 \abs{N_1-N_2}+2} \\
3 \frac{\sqrt{2\abs{N_1-N_2}\left[N_1 N_2 - \abs{N_1-N_2}
-1\right]}}{
 \abs{N_1-N_2}+2}  &
\frac{(N_1-N_2)^2}{4} + \frac{\abs{N_1-N_2}}{2} - 2
  + \frac{3}{2} \frac{N_1^2-N_2^2}{\abs{N_1-N_2}+2}
\end{pmatrix} \\ \times
\begin{pmatrix}
  \alpha_{\frac{\abs{N_1-N_2}}{2},\frac{\abs{N_1-N_2}}{2}} \\
  \alpha_{\frac{\abs{N_1-N_2}}{2}, \frac{\abs{N_1-N_2}}{2}+1}
\end{pmatrix}
= \lambda_{\frac{\abs{N_1-N_2}}{2}} \begin{pmatrix}
  \alpha_{\frac{\abs{N_1-N_2}}{2},\frac{\abs{N_1-N_2}}{2}} \\
  \alpha_{\frac{\abs{N_1-N_2}}{2}, \frac{\abs{N_1-N_2}}{2}+1}
\end{pmatrix},
  j = \frac{\abs{N_1-N_2}}{2} \neq 0, \min(N_1,N_2) \neq 1,
\end{aligned} \\
\begin{aligned}
\begin{pmatrix}
\tfrac{(N_1-N_2)^2}{4} + \tfrac{\abs{N_1-N_2}}{2} - 2 + 3
\sgn(N_1-N_2) \tfrac{N_1+N_2}{\abs{N_1-N_2}+2}
\end{pmatrix}
\begin{pmatrix}
\alpha_{\frac{\abs{N_1-N_2}}{2} \frac{\abs{N_1-N_2}}{2}}
\end{pmatrix}
\\ = \lambda_{\frac{\abs{N_1-N_2}}{2}}
\begin{pmatrix}
\alpha_{\frac{\abs{N_1-N_2}}{2} \frac{\abs{N_1-N_2}}{2}}
\end{pmatrix},
j = \frac{\abs{N_1-N_2}}{2}, \min(N_1,N_2)=1, N_1\neq N_2,
\end{aligned} \\
\begin{aligned}
\begin{pmatrix}
\frac{(N_1-N_2)^2}{4} - \frac{\abs{N_1-N_2}}{2} - 2 + \frac{3}{2}
(N_1+N_2) \sgn(N_1-N_2)
\end{pmatrix}
\begin{pmatrix} \alpha_{\frac{\abs{N_1-N_2}}{2}-1,\frac{\abs{N_1-N_2}}{2}}
\end{pmatrix} \\
= \lambda_{\frac{\abs{N_1-N_2}}{2}-1}
\begin{pmatrix} \alpha_{\frac{\abs{N_1-N_2}}{2}-1,\frac{\abs{N_1-N_2}}{2}}
\end{pmatrix}, \qquad
j = \frac{\abs{N_1-N_2}}{2}-1 \geq 0,
\end{aligned} \\
\begin{pmatrix} -2 \end{pmatrix} \begin{pmatrix} \alpha_{0,1,0} \end{pmatrix}
= \lambda_0 \begin{pmatrix} \alpha_{0,1,0} \end{pmatrix}, \qquad
N_1=N_2, j=0, \\
\begin{aligned}
\begin{pmatrix}
\frac{(N_1+N_2)^2}{4} - \frac{N_1+N_2}{2} - 2 - \frac{3}{2}
\frac{N_1^2-N_2^2}{N_1+N_2-2} &
3 \frac{\sqrt{2 (N_1-1) (N_2-1) (N_1+N_2)}}{N_1+N_2-2} \\
3 \frac{\sqrt{2 (N_1-1) (N_2-1) (N_1+N_2)}}{N_1+N_2-2} &
\frac{(N_1+N_2)^2}{4} - \frac{N_1+N_2}{2} - 2 + 3
\frac{N_1-N_2}{N_1+N_2-2}
\end{pmatrix}
\begin{pmatrix}
\alpha_{\frac{N_1+N_2}{2}-1,\frac{N_1+N_2}{2}-2} \\
\alpha_{\frac{N_1+N_2}{2}-1,\frac{N_1+N_2}{2}-1}
\end{pmatrix} \\
= \lambda_{\frac{N_1+N_2}{2}-1}
\begin{pmatrix}
\alpha_{\frac{N_1+N_2}{2}-1,\frac{N_1+N_2}{2}-2} \\
\alpha_{\frac{N_1+N_2}{2}-1,\frac{N_1+N_2}{2}-1}
\end{pmatrix}, \qquad j = \frac{N_1+N_2}{2}-1\geq 0,
\end{aligned} \\
\begin{pmatrix}
\frac{(N_1+N_2)^2}{4} + \frac{N_1+N_2}{2} - 2 - \frac{3}{2}
(N_1-N_2)
\end{pmatrix}
\begin{pmatrix} \alpha_{\frac{N_1+N_2}{2},\frac{N_1+N_2}{2}-1}
\end{pmatrix}
= \lambda_{\frac{N_1+N_2}{2}}
\begin{pmatrix} \alpha_{\frac{N_1+N_2}{2},\frac{N_1+N_2}{2}-1}
\end{pmatrix}, \qquad j = \frac{N_1+N_2}{2}.
\end{gather}
\end{subequations}%

For the generic ($\tfrac{\abs{N_1-N_2}}{2}+1\leq j \leq
\tfrac{N_1+N_2}{2}-2$) problem, the eigenvalues $\lambda$ are given
by
\begin{equation}
\lambda = (j+2)(j-1) + \tilde{\lambda},
\end{equation}
where
\begin{equation} \label{lambda}
\tilde{\lambda} = \sqrt{6} \sqrt{N_1^2+N_2^2-2 j(j+1)}
 \cos \left[\frac{\pi (2 n+1)}{3}
    - \frac{1}{3} \cos^{-1} \sqrt{\frac{27}{8}} \frac{N_1^2-N_2^2}{[
           N_1^2+N_2^2-2j(j+1)]^{3/2}} \right], n =0,\pm 1,
\end{equation}
are the roots of
\begin{equation}
\tilde{\lambda}^3 - 9\left[\frac{N_1^2+N_2^2}{2} -
j(j+1)\right]\tilde{\lambda} +\frac{27}{4} (N_1^2-N_2^2) = 0.
\end{equation}
The eigenvectors---for $N_1\neq N_2$---are then given by
\begin{subequations}
\begin{gather}
\begin{split}
4 \frac{[N_1^2-N_2^2-\frac{4}{3} \tilde{\lambda}(j+1)]
\sqrt{[j^2-\frac{(N-1-N_2)^2}{4}][\frac{(N_1+N_2)^2}{4}-j^2]}}{
    \sqrt{j(2j+1) \kappa}} {\boldsymbol{Y}}_{j,j-1,m}
+ \frac{[N_1^2-N_2^2-\frac{4}{3} \tilde{\lambda}(j+1)]
 [N_1^2-N_2^2+\frac{4}{3} \tilde{\lambda} j]}{\sqrt{j(j+1)\kappa}}
 {\boldsymbol{Y}}_{jjm} \\
- 4 \frac{[N_1^2-N_2^2+\frac{4}{3} \tilde{\lambda} j]
\sqrt{[(j+1)^2-\frac{(N-1-N_2)^2}{4}][\frac{(N_1+N_2)^2}{4}-(j+1)^2]}}{
    \sqrt{(j+1)(2j+1) \kappa}} {\boldsymbol{Y}}_{j,j+1,m},
\end{split} \\ \begin{split}
\kappa \equiv \left[\tfrac{256}{9}(N_1^2+N_2^2)j(j+1)
 - \tfrac{512}{9}j^2(j+1)^2 - \tfrac{16}{3}(N^2_1-N^2_2)^2\right]
      \tilde{\lambda}^2
- \left[\tfrac{320}{3}(N_1^2-N_2^2)j(j+1) -
\tfrac{64}{3}(N_1^4-N_2^4)\right]
     \tilde{\lambda}
\\
-16 \left[N_1^2-N_2^2\right]^2
\left[j^2+j+3-\tfrac{N_1^2+N_2^2}{2}\right].
\end{split}
\end{gather}
\end{subequations}
As checks, note that the $\tilde{\lambda}$'s sum to zero, which
agrees with the trace of the matrix of the eigenvalue problem (after
subtracting $(j+2)(j-1) \one$).  Also, the product of the
$\tilde{\lambda}$'s can be evaluated using
\begin{equation}
\prod_{n=-1}^1 \cos \left(\tfrac{2 \pi n}{3} + \tfrac{x}{3}\right) =
\prod_{n=-1}^1 \left(\cos \tfrac{2\pi n}{3} \cos \tfrac{x}{3}
  - \sin \tfrac{2\pi n}{3} \sin \tfrac{x}{3} \right)
= \cos^3 \tfrac{x}{3} - \tfrac{3}{4} \cos \tfrac{x}{3} =
\tfrac{1}{4} \cos x.
\end{equation}
Thus, the product of the $\tilde{\lambda}$'s is
\begin{equation}
\prod_{n=-1}^1 \tilde{\lambda} = 6^{3/2} (N_1^2+N_2^2-2j(j+1)^{3/2}
  \cos \left[ \pi - \cos^{-1} \sqrt{\frac{27}{8}} \frac{N_1^2-N_2^2}{[
           N_1^2+N_2^2-2j(j+1)]^{3/2}} \right]
= - \frac{27}{4} (N_1^2-N_2^2).
\end{equation}
which agrees with the determinant of the matrix.

When $j=\frac{N_1+N_2}{2}-1$, the eigenvectors and eigenvalues
$\tilde{\lambda}$ of the relevant matrix
\begin{equation}
\begin{pmatrix}
- \frac{3}{2} \frac{N_1^2-N_2^2}{N_1+N_2-2} &
3 \frac{\sqrt{2 (N_1-1) (N_2-1) (N_1+N_2)}}{N_1+N_2-2} \\
3 \frac{\sqrt{2 (N_1-1) (N_2-1) (N_1+N_2)}}{N_1+N_2-2} & + 3
\frac{N_1-N_2}{N_1+N_2-2}
\end{pmatrix},
\end{equation}
are
\begin{subequations}
\begin{gather}
\begin{alignat}{2}
&\sqrt{\frac{1}{2} \pm \frac{\upsilon}{2 \nu}}
      {\boldsymbol{Y}}_{\frac{N_1+N_2}{2}-1,\frac{N_1+N_2}{2}-2, m}
\pm \sqrt{\frac{1}{2} \mp \frac{\upsilon}{2 \nu}}
   {\boldsymbol{Y}}_{\frac{N_1+N_2}{2}-1,\frac{N_1+N_2}{2}-1, m}, & \qquad
\tilde{\lambda} &= -\frac{3}{2}(N_1-N_2) \pm \nu,
\end{alignat} \\
\begin{align}
\upsilon &\equiv -\frac{3}{4}
\frac{(N_1-N_2)(N_1+N_2+2)}{N_1+N_2-2}, & \nu &\equiv \frac{3}{4}
\sqrt{(N_1^2-N_2^2)^2 + 8 (N_1+N_2)}.
\end{align}
\end{gather}
\end{subequations}

For $N_1=N_2=N$, there is an enormous simplification of the generic
problem. We find the normalized eigenvectors
\begin{subequations}
\begin{alignat}{2}
-\sqrt{\frac{j[N^2-(j+1)^2]}{(2j+1)[N^2-j(j+1)]}}
{\boldsymbol{Y}}_{j,j-1,m}
 + \sqrt{\frac{(j+1)[N^2-j^2]}{(2j+1)[N^2-j(j+1)]}} {\boldsymbol{Y}}_{j,j+1,m},
 && \qquad \tilde{\lambda} &= 0, \\
\sqrt{\frac{(j+1)[N^2-j^2]}{2(2j+1)[N^2-j(j+1)]}}
  {\boldsymbol{Y}}_{j,j-1,m}
\pm \frac{1}{\sqrt{2}} {\boldsymbol{Y}}_{jjm} +
\sqrt{\frac{j[N^2-(j+1)^2]}{2(2j+1)[N^2-j(j+1)]}}
  {\boldsymbol{Y}}_{j,j+1,m},
&& \qquad \tilde{\lambda} &= \pm 3 \sqrt{N^2-j(j+1)}, \intertext{for
$1\leq j \leq \frac{N_1+N_2}{2}-2$, and}
\boldsymbol{Y}_{010}, && \qquad \tilde{\lambda} &= 0, \\
\frac{1}{\sqrt{2}} \boldsymbol{Y}_{N-1,N-2,m} \pm \frac{1}{\sqrt{2}}
\boldsymbol{Y}_{N-1,N-1,m}, &&
\qquad \tilde{\lambda} &= \pm 3 \sqrt{N}, N \neq 1, \\
\boldsymbol{Y}_{N,N-1,m}, && \qquad \tilde{\lambda} &= 0.
\end{alignat}
\end{subequations}

\section{Finding the masses of $\chi$} \label{sec:chimass}
Below we give the details of finding the masses for the fermionic
off-diagonal fluctuations $\chi$. This part of the action (\ref{eqn:
S2}) is (after integration by parts) \BE\label{eqn:
SFod}(S_\text{fermion})_\text{o.d.}=2\Tr\left(i\chi^\dagger\dot{\chi}+
\frac{1}{3}\chi^\dagger\gamma^i(\chi
J^i_{(2)}+J^i_{(1)}\chi)-\frac{x^a}{\mu}\chi^\dagger\gamma^a\chi
-i\frac{1}{4}\chi^\dagger\gamma^{123}\chi\right), \EE which gives the Dirac
equation \BE \left(i\partial_t-\frac{x^b}{\mu}\gamma^b-
i\frac{1}{4}\gamma^{123}+\frac{1}{3}\gamma^j
\anti{J^j}{\cdot}\right) \chi=0.\EE Squaring the Dirac equation,
\ie\ acting on the left with the conjugate Dirac operator
$\left(-i\partial_t-\frac{x^a}{\mu}\gamma^a-i\frac{1}{4}\gamma^{123}+\frac{1}{3}\gamma^i
\anti{J^i}{\cdot}\right)$ gives the ``Klein-Gordon'' equation \BE
\partial_t^2\chi
+i\frac{v^b}{\mu}\gamma^b\chi
+\frac{x^bx^b}{\mu^2}\chi
+\left(\frac{1}{4}\right)^2\chi
-i\frac{1}{6}\gamma^{123}\gamma^i\anti{J^i}{\chi}
+\left(\frac{1}{3}\right)^2\gamma^i\gamma^j\anti{J^i}{\anti{J^j}{\chi}}=0.
\EE
Upon setting $v^b=v\delta^{b9}$ the term
$i\frac{v^b}{\mu}\gamma^b\chi$ becomes $i\frac{v}{\mu}\gamma^9\chi$.
Since $\gamma^9$ commutes with the other gamma matrices, \ie,
$\gamma^{123}\gamma^i$ and $\gamma^i\gamma^j$, in the e.o.m., we can
first diagonalize w.r.t. $\gamma^9$. Upon the projection
\BE
\chi_\pm\equiv\frac{1\pm\gamma^9}{2}\chi,
\EE
the e.o.m. separates
into $\pm$ parts
\BE
\partial_t^2\chi_\pm\pm
i\frac{v}{\mu}\chi_\pm+\frac{x^bx^b}{\mu^2}\chi_\pm+\left(\frac{1}{4}\right)^2\chi_\pm
-i\frac{1}{6}\gamma^{123}\gamma^i \anti{J^i}{\chi_\pm}
+\left(\frac{1}{3}\right)^2\gamma^i\gamma^j\anti{J^i}{\anti{J^j}{\chi_\pm}}=0.
\EE
We see that the difference between the $+$ and $-$ components of
the e.o.m.\ is just $\chi_+\to\chi_-$ and $v\to -v$. Hence in the
following let's first concentrate on the $\chi_+$ equation and
suppress the $+$ subscript. Readily seen, solutions to the
eigenproblem
\BE\label{eqn: fermion-eigen}
i\gamma^{123}\gamma^i \anti{J^i}{\chi}=\lambda\chi,
\EE
(which implies that
$\gamma^i\gamma^j\anti{J^i}{\anti{J^j}{\chi}}=\lambda^2\chi$) diagonalize the
e.o.m., giving the mass squared for $\chi_+$ (after Wick rotation
$v\to -iv$)\BE\label{eqn: mass of chiplus}
m_{\chi_+}^2=\frac{v}{\mu}+\frac{x^ax^a}{\mu^2}+\left(\frac{\lambda}{3}-\frac{1}{4}\right)^2,\EE
and similarly the mass-squared for $\chi_-$
\BE\label{eqn: mass of chiminus}
m_{\chi_+}^2 =
-\frac{v}{\mu}
+\frac{x^a x^a}{\mu^2}+\left(\frac{\lambda}{3}-\frac{1}{4}\right)^2.
\EE
We first look at the eigenproblem (\ref{eqn: fermion-eigen}) for
$\chi_+$. Adopting the gamma matrix representation of \cite{GSW}
where
\BE
\gamma^9=\begin{pmatrix}I_{8\times 8}&0\\
0 &-I_{8\times 8}\end{pmatrix}, \qquad
\gamma^i=\begin{pmatrix}0&\tilde{\gamma}^i\\
(\tilde{\gamma}^i)^T&0\end{pmatrix},
\EE
(with
$\tilde{\gamma}^i$'s being $8\times 8$ matrices; more specifically,
$\tilde{\gamma}^1=i\tau_2\otimes i\tau_2\otimes i\tau_2$,
$\tilde{\gamma}^2=1\otimes \tau_1\otimes i\tau_2$, and
$\tilde{\gamma}^3=1\otimes\tau_3\otimes i\tau_2$, with
$\tau_1,\tau_2,\tau_3$ being the standard Pauli matrices), we see
that
$\chi_+$ is of the form
\BE
\chi_+=\begin{pmatrix}\theta_+\\ 0\end{pmatrix},
\EE
where $\theta_+$ is a $8$-component spinor.
Then eqn.~(\ref{eqn: fermion-eigen}) becomes
\BE\label{eqn: theta-plus}
1\otimes i\tau_2\otimes 1 \, \anti{J^1}{\theta_+}
-i\tau_2\otimes\tau_1\otimes 1\, \anti{J^2}{\theta_+}
-i\tau_2\otimes \tau_3\otimes 1\, \anti{J^3}{\theta_+}
= i \lambda\theta_+.
\EE
Similarly
$\chi_-$ is of the
form
\BE
\chi_-= \begin{pmatrix} 0\\ \theta_-\end{pmatrix},
\EE
and the eigen-equation for $\chi_-$ in
terms of $\theta_-$ is the same as eqn.~(\ref{eqn: theta-plus}) with
$\theta_+\to\theta_-$. Hence we see that $\chi_+$ and $\chi_-$ have
the same eigenvalue $\lambda$. To solve eqn.~(\ref{eqn: theta-plus})
we do the projection
\BE
\theta_+=\theta_1+\theta_2,\text{with }
\theta_1\equiv \frac{1+\tau_2}{2}\otimes 1\otimes 1\ \theta_+,\
\theta_2\equiv \frac{1-\tau_2}{2}\otimes 1\otimes 1\ \theta_+,
\EE
which diagonalizes the first gamma matrix in the direct-product of
three of them in eqn.~(\ref{eqn: theta-plus}) (the third gamma matrix
is already diagonalized automatically). Hence we can write
$\theta_1$ as a two-component spinor acted upon by the second gamma
matrix and the eigen-equation becomes
\BE\label{eqn: theta1}
\anti{\tau_2 J^1-\tau_1 J^2-\tau_3J^3}{\theta_1}
=\lambda_1\theta_1,
\EE
(where we
have added the subscript $1$ to $\lambda$) and note that for each
value of $\lambda_1$ there is a degeneracy of two $\theta_1$'s
(coming from the third gamma matrix, recalling that $\theta_1$ has
four real degrees of freedom). Similarly, the equation for
$\theta_2$ reads
\BE\label{eqn: theta2}
\anti{\tau_2 J^1+\tau_1 J^2+\tau_3J^3}{\theta_2}
=\lambda_2\theta_2,
\EE
(where we have added the subscript $2$ to $\lambda$)
with a two-fold degeneracy for each
value of $\lambda_2$. Below let us first solve the equation for
$\theta_1$.

Denote the matrix $(\tau_2 J^1-\tau_1 J^2-\tau_3J^3)$ as $\Delta_1$
and act it on $\theta_1$ twice, we get
\BE
\anti{\Delta_1}{\anti{\Delta_1}{\theta_1}}=\lambda_1^2\theta_1,
\EE whose l.h.s.\
after some algebra can be written as
\BE\label{eqn: lhs of Delta square}
2(\Lambda_{(1)}+\Lambda_{(2)})\theta_1-\com{J^i}{\com{J^i}{\theta_1}}
+\tau_2\com{J^1}{\theta_1}
-\tau_1\com{J^2}{\theta_1}-\tau_3\com{J^3}{\theta_1}.
\EE
Then we write
\BE
\theta_1=\begin{pmatrix}\beta\\ \eta\end{pmatrix},
\EE
expand $\beta,\eta$
\BE \beta=\sum_{l=\frac{\left\lvert
N_1-N_2\right\rvert}{2}}^{\frac{N_1+N_2}{2}-1}\sum_{m=-l}^l\beta_{lm}Y_{lm},\
\eta=\sum_{l=\frac{\left\lvert
N_1-N_2\right\rvert}{2}}^{\frac{N_1+N_2}{2}-1}\sum_{m=-l}^l\eta_{lm}Y_{lm},
\EE
and plug them into eqn.~(\ref{eqn: lhs of Delta square}). Solving
the resulting equation we find the eigenvectors for the operator
$\anti{\Delta_1}{\anti{\Delta_1}{\cdot}}$, which we summarize below:

$l=\frac{\left\lvert
N_1-N_2\right\rvert}{2},\dots,\frac{N_1+N_2}{2}-1$  and for any given
$l$ there are two cases:
\begin{caselist}
\item \label{caseA}
$\lambda_1^2=\frac{N_1^2+N_2^2}{2}-(l+1)^2$, which has
a degeneracy of $2l+2$, with $2l$ states given by
\BE
\begin{pmatrix}i\sqrt{\frac{l+m+1}{l-m}}Y_{lm}\\ Y_{l,m+1}\end{pmatrix}
\text{ for } m=-l,\dots,l-1,
\EE and the other $2$ states being
\BE
\begin{pmatrix}0\\ Y_{l,-l}\end{pmatrix} \text{ and }
\begin{pmatrix}Y_{ll}\\ 0\end{pmatrix}.
\EE

\item \label{caseB} $\lambda_1^2=\frac{N_1^2+N_2^2}{2}-l^2$, which has a
degeneracy $2l$, with the states given by
\BE
\begin{pmatrix}-i\sqrt{\frac{l-m}{l+m+1}}Y_{lm}\\ Y_{l,m+1}\end{pmatrix}
\text{ for } m=-l,\dots,l-1.
\EE
\end{caselist}
One can check that the total number of states in
cases~\ref{caseA} and~\ref{caseB} is equal to $2N_1N_2$ as expected.

Next we solve for the eigenvectors of the operator $\anti{\Delta_1}{\cdot}$ by
making linear combinations of the those of
$\anti{\Delta_1}{\anti{\Delta_1}{\cdot}}$
found above. To do this we have to make extensive use of the formula
(\ref{antiJY}). We find
\begin{itemize}
\item Take the extremal value $l=\frac{N_1+N_2}{2}-1$ in the
case~\ref{caseA} above, all the $2l+2=N_1+N_2$ eigenvectors of
$\anti{\Delta_1}{\anti{\Delta_1}{\cdot}}$ worked out above are automatically
eigenvectors of $\anti{\Delta_1}{\cdot}$, with eigenvalue being
$\lambda_1=\frac{N_2-N_1}{2}$.

\item Take the extremal value $l=\frac{\left\lvert
N_1-N_2\right\rvert}{2}$ in case~\ref{caseB}, all the $2l=\left\lvert
N_1-N_2\right\rvert$ eigenvectors of $\anti{\Delta_1}{\anti{\Delta_1}{\cdot}}$
worked
out above are automatically eigenvectors of $\anti{\Delta_1}{\cdot}$, with
eigenvalue being
$\lambda_1=\left(\frac{N_1+N_2}{2}\right)\left(\frac{N_1-N_2}{\left\lvert
N_1-N_2\right\rvert}\right)$.

\item For generic values of $l$, one has to choose a state from
case~\ref{caseA} and a state from
case~\ref{caseB} and linearly combine them. The result
is:  $l=\frac{\left\lvert
N_1-N_2\right\rvert}{2},\dots,\frac{N_1+N_2}{2}-2$, for any given $l$,
the eigenvalues of $\anti{\Delta_1}{\cdot}$ are
$\lambda_1=\pm\sqrt{\frac{N_1^2+N_2^2}{2}-(l+1)^2}$ with a
degeneracy $2l+2$. We omit the expressions of the eigenvectors here
since those are long and won't be needed anyway.
\end{itemize}

As one can check, the total number of the above eigenvectors for
$\anti{\Delta_1}{\cdot}$ is \BE N_1+N_2+\left\lvert
N_1-N_2\right\rvert+\sum_{l=\frac{\left\lvert
N_1-N_2\right\rvert}{2}}^{\frac{N_1+N_2}{2}-2}2(2l+2)=2N_1N_2,\EE as
expected. This completes our solving the eigenproblem (\ref{eqn:
theta1}).

The eigen-equation (\ref{eqn: theta2}) which we write as
$\anti{\Delta_2}{\theta_2}=\lambda_2\theta_2$ with $\Delta_2\equiv (\tau_2
J^1+\tau_1 J^2+\tau_3J^3)$ is now easy to solve. One can readily
check that if $\theta_1=\begin{spmatrix}\beta\\
\eta\end{spmatrix}$ satisfies
$\{\Delta_1,\theta_1\}=\lambda_1\theta_1$, then
$\theta_2=\begin{spmatrix}-\eta\\\beta\end{spmatrix}$ satisfies
$\{\Delta_2,\theta_2\}=\lambda_2\theta_2$ with
$\lambda_2=\lambda_1$, \ie\ $\Delta_2$ has the same eigenvalues and
degeneracies as $\Delta_1$ does.

Let us summarize the off-diagonal fermionic fluctuation $\chi$'s
mass spectrum.
The mass spectrum of $\chi_+$ (which has a total number of $8N_1N_2$
real d.o.f.'s) is 
\BE\label{eqn: fermion mass spectrum}
m_{\chi_+}=\sqrt{\frac{v}{\mu}+\frac{x^ax^a}{\mu^2}+\left(\frac{\lambda}{3}-\frac{1}{4}\right)^2},
\EE
with
\begin{equation}\label{eqn: fermion mass spectrum detail}
\begin{aligned}
&&\lambda&=\frac{N_2-N_1}{2},& \text{degeneracy: }& 4(N_1+N_2),\\
&&\lambda&=\frac{N_1+N_2}{2}\left(\frac{N_1-N_2}{\left\lvert N_1-N_2
\right\rvert}\right),& \text{degeneracy: }& 4\left\lvert N_1-N_2
\right\rvert,\\
l&=\frac{\left\lvert N_1-N_2
\right\rvert}{2},\dots,\frac{N_1+N_2}{2}-2, &
\lambda&=\pm \sqrt{\frac{N_1^2+N_2^2}{2}-(l+1)^2}, &
\text{degeneracy: }&8l+8, \text{ for each sign}.
\end{aligned}
\end{equation}
The mass spectrum of $\chi_-$ (which also has a total number of
$8N_1N_2$ real d.o.f.'s) is obtained from that of $\chi_+$ by simply
changing $v$ to $-v$.

\end{document}